\documentclass[twocolumn,amsmath,amssymb,superscriptaddress,aps,pra,a4paper,floatfix]{revtex4-2}

\usepackage{graphicx}
\usepackage[paperwidth=212mm,paperheight=297mm,centering,hmargin=1.65cm,vmargin=2.6cm]{geometry}
\usepackage{color} 
\usepackage{dcolumn}
\usepackage{bm}

\usepackage{bbm}
\usepackage{dsfont}
\usepackage{orcidlink}

\usepackage{booktabs}
\usepackage{physics}
\AtBeginDocument{\RenewCommandCopy\qty\SI} 
\usepackage[version=3]{mhchem}
\usepackage{textcomp}

\usepackage{float}
\usepackage[separate-uncertainty = true,multi-part-units=single]{siunitx}

\DeclareSIUnit{\molar}{M}
\DeclareSIUnit{\counts}{cnt}

\usepackage[T1]{fontenc}
\newcommand\thefont{\expandafter\string\the\font}

\usepackage{subfigure}
\usepackage{subcaption}
\usepackage{jabbrv}
\usepackage{chemformula}
\usepackage{natbib}
\usepackage{multirow}
\usepackage{dsfont}
\usepackage{comment}
\usepackage{makecell}
\usepackage{tabularx}
\usepackage{hyperref}
\usepackage[capitalise]{cleveref}

\usepackage{braket}

\RenewCommandCopy\qty\SI

\usepackage[dvipsnames]{xcolor} 
\hypersetup{
    colorlinks = true, 
    urlcolor   = BlueViolet, 
    linkcolor  = BlueViolet, 
    citecolor  = BlueViolet    
}

\usepackage{xr}
\makeatletter
\newcommand*{\addFileDependency}[1]{
  \typeout{(#1)}
  \@addtofilelist{#1}
  \IfFileExists{#1}{}{\typeout{No file #1.}}
}
\makeatother

\makeatletter
\def\maketitle{
\@author@finish
\title@column\titleblock@produce
\suppressfloats[t]}
\makeatother

\newcommand{\TwoEightSi}{\ensuremath{^{28}\text{Si}} }

\newcommand{\triplet}{1s:\ensuremath{\mathrm{^{3}T_2}}} 
\newcommand{\singlet}{1s:\ensuremath{\mathrm{^{1}T_2}}} 
\newcommand{\ground}{1s:\ensuremath{\mathrm{^{1}A_1}}} 

\newcommand{\stateE}{1s:\ensuremath{\mathrm{E}}}
\newcommand{\stateT}{1s:\ensuremath{\mathrm{T_2}}} 
\newcommand{\stateA}{1s:\ensuremath{\mathrm{A_1}}}

\newcommand{\Als}{Al\ensuremath{_\text{s}}}
\newcommand{\Ali}{Al\ensuremath{_\text{i}}}
\newcommand{\Aliplus}{Al\ensuremath{_\text{i}^{+}}}
\newcommand{\Aliplusplus}{Al\ensuremath{_\text{i}^{++}}}

\begin{document}
\title{Optically Resolved Excited State Hyperfine Structure of a Silicon Colour Centre in the Telecom Bands}

\author{A.~Woolverton \orcidlink{0009-0000-4276-0030}}
\affiliation{Department of Physics, Simon Fraser University, Burnaby, British Columbia, Canada}
\affiliation{Photonic Inc., Coquitlam, British Columbia, Canada}

\author{N.~Brunelle \orcidlink{0000-0002-0224-5613}}
\affiliation{Department of Physics, Simon Fraser University, Burnaby, British Columbia, Canada}
\affiliation{Photonic Inc., Coquitlam, British Columbia, Canada}

\author{M.~Keshavarz \orcidlink{0000-0002-3772-3304}}
\affiliation{Department of Physics, Simon Fraser University, Burnaby, British Columbia, Canada}
\affiliation{Photonic Inc., Coquitlam, British Columbia, Canada}

\author{L.~Bergeron \orcidlink{0000-0002-3318-6345}}
\affiliation{Department of Physics, Simon Fraser University, Burnaby, British Columbia, Canada}
\affiliation{Photonic Inc., Coquitlam, British Columbia, Canada}

\author{E.~R.~MacQuarrie \orcidlink{0000-0003-3097-5549}}
\affiliation{Department of Physics, Simon Fraser University, Burnaby, British Columbia, Canada}
\affiliation{Photonic Inc., Coquitlam, British Columbia, Canada}

\author{Y.~Ackermann}
\affiliation{Department of Physics, Simon Fraser University, Burnaby, British Columbia, Canada}
\affiliation{Photonic Inc., Coquitlam, British Columbia, Canada}

\author{M.~Gascoine \orcidlink{0009-0002-0185-166X}}
\affiliation{Department of Physics, Simon Fraser University, Burnaby, British Columbia, Canada}
\affiliation{Photonic Inc., Coquitlam, British Columbia, Canada}

\author{N.~Abrosimov \orcidlink{0000-0003-3271-9602}}
\affiliation{Leibniz-Institut für Kristallzüchtung, Berlin, Germany}

\author{S.~Simmons \orcidlink{0000-0002-5824-736X}}
\affiliation{Department of Physics, Simon Fraser University, Burnaby, British Columbia, Canada}
\affiliation{Photonic Inc., Coquitlam, British Columbia, Canada}

\author{D.~B.~Higginbottom \orcidlink{0000-0002-9825-9840}} 
\email{daniel_higginbottom@sfu.ca}
\affiliation{Department of Physics, Simon Fraser University, Burnaby, British Columbia, Canada}
\affiliation{Photonic Inc., Coquitlam, British Columbia, Canada}

\author{M.~L.~W.~Thewalt \orcidlink{0000-0002-5806-0618}}
\affiliation{Department of Physics, Simon Fraser University, Burnaby, British Columbia, Canada}

\begin{abstract}
Nuclear spin qubits in silicon offer exceptionally coherent quantum memory, and optically-interfaced spins are a promising platform for both quantum networking and distributed quantum computing. It has been proposed that emitters with diamagnetic ground states may permit an optical interface to nuclear spin memories via metastable, hyperfine-coupled excited states while suppressing key sources of decoherence. Until now, direct optical observation of suitable transitions in silicon colour centres has remained elusive. Here we characterize the singly-ionized interstitial aluminum donor (\Aliplus) in isotopically purified \TwoEightSi{} and find several novel features of this little-studied defect. We measure bright emission and strong optical transitions and, in contrast to previous studies of this centre, attribute its emission to an exchange-split spin triplet and singlet level of the lowest-energy \stateT{} excited state. We measure the excited-state lifetimes and, as a consequence of its narrow emission linewidth, observe the fine and hyperfine structure of the long-lived triplet state. This constitutes the first measurement of an optically-resolved hyperfine structure in the excited state of a telecommunications-band silicon colour centre. 
\end{abstract}

\date{\today}
\maketitle

\section{Introduction}
Silicon colour centres are a rapidly developing platform for quantum computing and networking. They can provide an interface between long-lived nuclear spin qubits and telecommunications-band photons. Isotopically-purified \ce{^{28}Si} provides a nearly spin-free host and is capable of supporting nuclear spin coherence up to several hours \cite{Tyryshkin2012,Steger2012a,Saeedi2013}, and silicon emitters can be directly integrated into industry-standard silicon photonic circuits \cite{Buckley2017}. Cavity-integrated silicon colour centres have shown dramatic Purcell enhancement \cite{Gritsch2023Purcell, Johnston_2024_CavityCoupledTcentre}, high emission efficiency \cite{Lefaucher2025BrightMicrocavity}, indistinguishable emission \cite{Komza2024IndistinguishablePhotonics,Afzal2024}, electrically-triggered emission \cite{dobinson2025Electrically,day2025Probing}, and nuclear spin coherence times above \qty{200}{\ms} \cite{Song2025LongLived,Afzal2024}. These colour centres may function as a spin-photon interface if they are capable of entangling single photons with spins through spin-selective optical transitions. Long-lived nuclear spins, either native to the host material or the centre itself, function as `memory' qubits that are hyperfine-coupled to an optically-resolved electron spin `communication' qubit \cite{Benjamin2006BrokeredComputation,Pompili2021a}.

Solid-state defects may be classed by the presence or absence of an electron spin in the ground state. Shallow defects with paramagnetic ground states, such as phosphorous donors \cite{Steger2012a}, are foundational candidates for silicon spin qubits \cite{Kane1998}. Alternatively, deep centres in silicon, including rare-earth ions like Er$\mathrm{^{3+}}$\cite{Yang2022}, donors such as Se$\mathrm{^+}$\cite{Deabreu2019}, and colour centres such as the T centre \cite{Bergeron:2020_PRX, Clear2024optical, Afzal2024}, are well-studied candidates for spin-photon interfaces. Silicon colour centres with diamagnetic ground states, including G \cite{ODonnell1982, Watkins1982, Beaufils2018, Cache2025, Cache2026OpticalSilicon}, C \cite{Udvarhelyi2021,Ishikawa2011OpticalSilicon,Wen2025OpticalSpin}, and interstitial carbon (C$_\mathrm{i}$) \cite{Deak2024QuantumBit,Jhuria2024ProgrammableQuantum}, have recently been proposed as alternative interfaces to nuclear memory qubits via their metastable excited states. These schemes may be advantageous for quantum memories since they remove electron-induced decoherence in the ground state.  Comparable diamagnetic diamond colour centres have demonstrated coherent state transfer and nuclear state readout through a metastable state \cite{Lee2013ReadoutAncilla}, coherent state transfer has been demonstrated between silicon nuclear spins and the metastable triplet state of the oxygen-vacancy centre \cite{Akhtar2012CoherentSilicon}, and coherent control of the electron spin has been demonstrated in ensembles of G centres \cite{Cache2026OpticalSilicon}. Furthermore, the hyperfine structure of the G centre has been observed by optically detected magnetic resonance (ODMR) \cite{ODonnell1982, Watkins1982, Cache2025} but not directly in photoluminescence (PL). Wen \textit{et al.} \cite{Wen2025OpticalSpin} observed the C centre's fine structure by ODMR while a hyperfine structure was noticeably absent. Similar structure for C$\mathrm{_i}$ is expected \cite{Deak2024QuantumBit}. 

Here, we report the first observation of optically resolved hyperfine structure within the excited state of a silicon colour centre with a diamagnetic ground state. We consider singly-ionized interstitial aluminum (\Aliplus) which emits in extremely low-loss telecommunications bands and find many potential advantages of this system. Early observations of \Aliplus{} measured its strongly-absorbing optical transitions \cite{ Latushko1982IRAluminum, Latushko1989IRSilicon}. Although these properties have been known for several decades, \Aliplus{} has not been considered as a quantum platform prior to this work. We perform optical absorption and photoluminescence spectroscopy on an \Aliplus{} ensemble in isotopically-enriched $^{28}$Si. We measure the \stateT{} singlet and triplet excited state lifetimes, and successfully resolve both fine and hyperfine structure within the spin-triplet. We find the triplet state is both long-lived and sufficiently sharp to resolve the hyperfine coupling to the aluminum nuclear spin. 

\section{Review} \label{sec: review} 
Substitutional aluminum in silicon (\Als) has a well-studied shallow acceptor level, but little information has been reported on the readily-formed and stable interstitial impurity (\Ali). Latushko \textit{et al.}\,\cite{ Latushko1982IRAluminum, Latushko1989IRSilicon, Latushko1989IdentificationSi:Al} observed the He$^{+}$-like donor absorption series of \Aliplus, referred to as the AM1 series, and reported an ionization energy near \qty{930}{\meV}. They observed photoluminescence at \qty{774.8(1)}{meV} (in the telecom L band) and \qty{796.4(1)}{meV} (in the C band) and attributed these to the transitions between a spin-orbit split \stateT{} state and the \stateA{} ground state \cite{Grimmeiss1982}.

In this work, we demonstrate that these transitions instead correspond to a spin-triplet (\triplet) and singlet (\singlet) originating from the two-electron exchange interaction, where the triplet state is lower in energy due to Hund's rule. Additional fine structure arises from spin–orbit interactions acting within the orbitally degenerate \stateT{} manifold~\cite{Bergman1988}. This hypothesis was constructed by drawing similarities to the observed donor series of neutral chalcogens \cite{Bergman1986ObservationSilicon, Bergman1988, Janzen1984, Grimmeiss1982, Steger2009, Swartz1980OpticalSilicon} and magnesium double-donors \cite{Abraham2018FurtherSilicon, Pavlov2019, Thilderkvist1994InterstitialSilicon}, although \Ali{} is more appropriately described as a triple donor due to its 3 valence electrons. In this case, the $3+$ charge state is inaccessible and resides deep within the valence band \cite{Brower1970ElectronSilicon}. Singlet-triplet splittings occur in the electronic excited states of neutral double donors, such as Mg$_{\text{i}}$ or the chalcogens, but are found in the $1+$ state for \Ali.

\begin{figure}[t]
    \centering
    \includegraphics{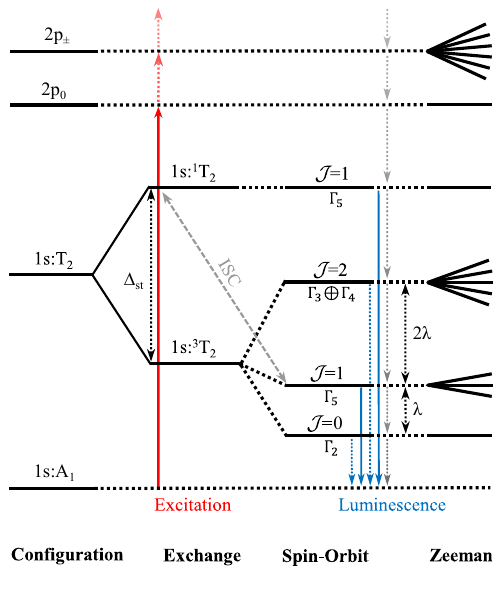}
    \caption{Proposed level structure of \Aliplus{} and PL excitation scheme. Each excited state includes an electron in the 1s:$\mathrm{A_1}$ ground level, and the 1s:$\mathrm{T_2}$ state splits under the exchange interaction. An emergent SOC separates the triplet state into effective $\mathcal{J}$ levels with degeneracy further lifted by the Zeeman interaction. The energy seperations are not to scale. The ISC allows for preferential population into the $\mathrm{\mathcal{J}=1}$ triplet level. Optical excitation to 2p$_0$, 2p$_\pm$, and above-band (red) excitation, used for our PL measurements, are followed by rapid decay to the \stateT{} manifold. Luminescence is observed from \singlet{} and \triplet{} (blue).}
    \label{fig: level structure}
\end{figure}

The proposed level structure of \Aliplus{} is shown in \cref{fig: level structure}. Within the two-electron ground state, each electron resides in the 1s:$\mathrm{A_1}$ orbital with $S=0$, and the hyperfine interaction vanishes. It may, however, be present within the excited states where $S\neq0$. In the first excited state, one electron is promoted into the \stateT{} manifold which is further split by the Coulombic repulsion and exchange interaction with the remaining \stateA{} electron. Within \triplet, an effective total angular momentum ($\mathcal{J}$) emerges from the underlying orbital and spin degeneracy, with levels separated by the spin-orbit interaction \cite{Peale1988ZeemanSilicon}. This emergent spin-orbit coupling (SOC) allows the $\mathrm{\Gamma_5}$ component of the triplet and singlet to mix through the inter-system crossing (ISC), relaxing the $\mathrm{\Delta S=0}$ spin selection rule and enabling \triplet$\mathrm{(\mathcal{J}=1)\Rightarrow}$ \ground{} luminescence.
 
\section{Donor Absorption Series, PL Spectrum, and Lifetimes}
We measure the He$^{+}$-like donor series (AM1 series) and excited state structure of an \Aliplus{} ensemble in isotopically-enriched \TwoEightSi{} by absorption and PL spectroscopy. Sample preparation details are included in \cref{sec: appendix sample prep}. The absorption spectrum of the AM1 series (SM \cref{fig: SM AM1 absorption} \cite{Supp_Mat}) shows a significant absorption depth of 10\% on \singlet{} (\qty{796.4}{\meV}) and 15\% on $\mathrm{2p_\pm}$ (\qty{946.6}{\meV}). In addition, we identify several previously unobserved excited states, up to $\mathrm{6p_\pm}$. We refine the precision of previously observed transition energies by an order of magnitude. \Cref{sec: appendix AM1 table} gives a complete table of all transitions observed in this work. 

\begin{figure}[t]
    \centering 
    \includegraphics[width=8.5cm]{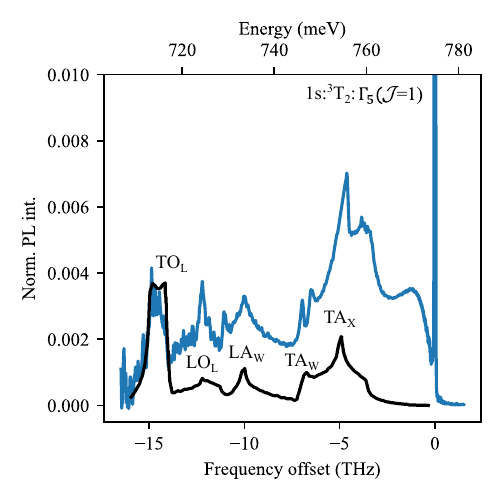}
    \caption{(Blue) Photoluminescence spectrum of \Aliplus{} at \qty{4.2}{\kelvin} under resonant excitation of $\mathrm{2p_\pm}$. The amplitude of the ZPL has been normalized to unity and truncated. The phonon side-band is compared to the silicon density of states (black) \cite{DeTomas2014ThermalModel}. We indicate the transverse optical (TO), transverse acoustic (TA), longitudinal optical (LO), and longitudinal acoustic (LA) phonon modes at high symmetry points within the first Brillouin zone \cite{Beaufils2018}.} 
    \label{fig: sideband spectrum}
\end{figure}

We then measure the PL spectrum at liquid helium temperature, shown in \cref{fig: sideband spectrum}, by resonantly exciting the \ground$\Leftrightarrow$$\mathrm{2p_\pm}$ transition. The PL spectrum is dominated by zero-phonon emission from \triplet{} to the ground state and its phonon sideband (PSB) which broadly matches the phonon density of states of silicon, shown for comparison. We calculate a Debye-Waller factor of \qty{47(1)}{\%} by numerical integration, comparing the ZPL area to the total spectrum.

Luminescence from \singlet{} is negligible below \qty{25}{K} as the population thermalizes to \triplet. Luminescence from \triplet{}\,$\Rightarrow$\,\ground{} is very bright and easily saturated, yet this transition is absent from the corresponding absorption spectrum due to its long radiative lifetime. We determine a splitting of $\Delta E_\mathrm{{st}} = \qty{21.26(1)}{\meV}$ from the \singlet{} and \triplet{} transition energies observed in these absorption and PL measurements, respectively. We determine the singlet and triplet state lifetimes of $\tau_\mathrm{s}=\qty{1.8(0.2)}{\us}$ and $\tau_\mathrm{t}=\qty{25.5(0.5)}{\ms}$ by measuring (1) their temperature-dependent luminescence transients under pulsed excitation (see \ref{sec: appendix lifetimes}), and (2) their temperature-dependent PL intensity ratio (see \ref{sec: appendix intensity ratio}).  

\section{Fine Structure} \label{sec: excited state}
\begin{figure}[b t]
    \centering
    \includegraphics[width = 8.5cm]{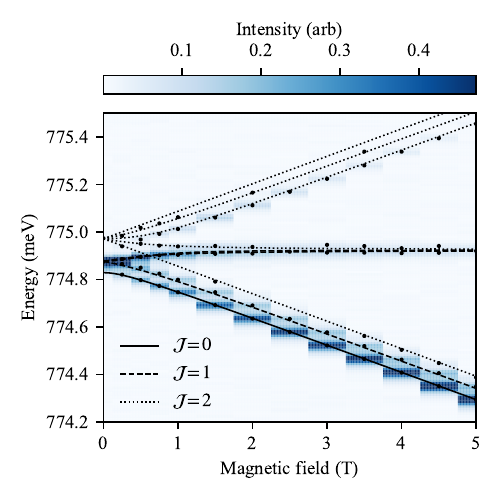}
    \caption{PL spectra of \triplet{} at \qty{4.2}{K} under above-band excitation and applied magnetic field, with $\mathrm{\hat{\mathbf{B}} \approx}$ [110]. Each dot represents an extracted peak position. The eigenenergies of each sublevel within $\mathcal{J}=0$, 1, and 2 are determined from the exact solutions of the Hamiltonian (SM  \cref{sec: sup Zeeman} \cite{Supp_Mat}) and shown as solid, dashed, and dotted lines.}  
     \label{fig: Zeeman eff}
\end{figure}

We observe fine structure splitting within \triplet{} under an applied magnetic field, shown in \cref{fig: Zeeman eff}.  We model this behaviour as a weak spin-orbit interaction using a fictitious angular momentum of $\mathcal{L}=1$ to reflect the triply degenerate orbital configuration \cite{Peale1988ZeemanSilicon, Abragam1951TheoryCrystals}. The spin and orbital degeneracies couple into total angular momentum $\mathcal{J}$, and the degeneracy is lifted through the effective SOC. Three components are expected with total effective momentum $\mathcal{J} \in \{0,1,2\}$, with corresponding $m_\mathcal{J}$ components that split in the presence of a magnetic field.

The electron Hamiltonian is given in \cref{eq: electron Zeeman Hamiltonian}. The first term represents an effective SOC with strength determined by $\lambda$. The remaining terms are the spin and orbital Zeeman effect where $\mu_\mathrm{B}$ is the Bohr magneton.
\begin{equation}
    \label{eq: electron Zeeman Hamiltonian} 
    \mathcal{H}_{\mathrm{e}}= \lambda(\mathbf{S}\cdot\boldsymbol{\mathcal{L}}) + 
    \mu_\mathrm{B} g_\mathrm{s}(\mathbf{S}\cdot\mathbf{B}) + \mu_\mathrm{B} g_\mathrm{\mathcal{L}}(\boldsymbol{\mathcal{L}}\cdot\mathbf{B})
\end{equation}
\noindent Following Peale \textit{et al}.\,\cite{Peale1988ZeemanSilicon}, we set $g_\mathrm{s}=2$ as both shallow single donors in silicon and the singly ionized donors Se$^+$ and Te$^+$ depart only slightly from the free-electron value. Similarly, the effective orbital Land\'e $g$-factor is $g_\mathcal{L}\approx0$ \cite{J.M.Luttinger1955MotionFields}. If the spin-orbit splitting is large compared to the Zeeman interaction, then only transitions to the $\Gamma_5(\mathcal{J}=1)$ component are observed. 

The SOC parameter $\lambda$ and unperturbed singlet energy relative to the ground state $E_\mathrm{t}$ are extracted from a least squares fit between the measured transition energies at \qty{4.2}{K} in \cref{fig: Zeeman eff} and the eigenenergies of \cref{eq: electron Zeeman Hamiltonian}. The transitions from the magnetic sublevels within the $\mathcal{J}=2$ level to \ground{} are less visible due to the selection rules of a magnetic dipole and instead thermalize to other sublevels. The fitted SOC parameter is $\lambda=$\qty{47.6(0.9)}{\micro eV} and unperturbed triplet state energy is $E_\mathrm{t}=$\qty{774.91(0.01)}{\meV}. 

Any orbital contributions arising from second-order perturbations and central-cell correction will cause an additional splitting within the \singlet{} state due to its angular degeneracy. No such splitting was observable within the linewidths of the measured absorption spectrum (SM \cref{fig: SM AM1 absorption}). 

\section{Hyperfine structure}
\begin{figure}[t]
    \centering 
    \includegraphics[width = 8.5cm]{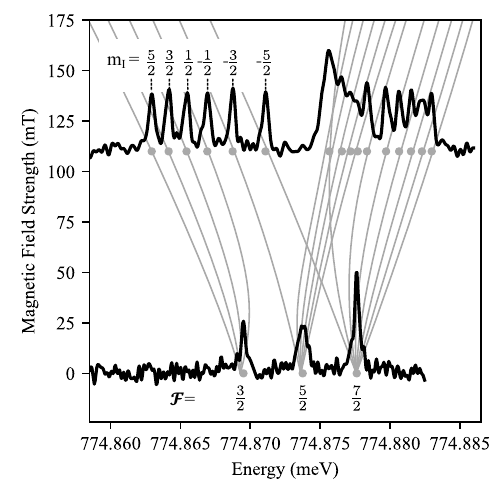}
    \caption{The optically resolved hyperfine structure within the \triplet$(\mathcal{J}=1)$\,$\Rightarrow$\,\ground{} transition is shown for 0\,mT (bottom) and 110\,mT (top) with a fit to the electronic spin Hamiltonian and isotropic hyperfine (grey). Each PL spectrum was measured at \qty{4.2}{\kelvin} using a resonant laser tuned to the \ground\,$\mathrm{\Rightarrow 2p_\pm}$ transition. The peak positions, shown as grey dots, are determined from a least-squares fit to multiple Gaussian-Lorentzian products with a shared background. The uncertainties of the transition energies are within the marker size.} 
    \label{fig: hyperfine spectrum}
\end{figure}

We observe three zero-field transitions within the $\mathcal{J}=1$ manifold, as shown in \cref{fig: hyperfine spectrum}, and label these levels with an effective total angular momentum $\mathcal{F}$ formed through the coupling of $\mathcal{J}$ and the Al nuclear spin with $\mathrm{I=5/2}$. SM \cref{sec: sup Zero field hyperfine early} \cite{Supp_Mat} presents an additional measurement made at \qty{1}{K} with an average linewidth comparable to the apparatus resolution (0.002\,$\mathrm{cm^{-1}}$).

We perform the same measurement at a magnetic field of $\mathrm{B_0=\qty{109.9(0.1)}{\mT}}$ and fit the Hamiltonian from \cref{eq: electron Zeeman Hamiltonian} with additional nuclear Zeeman and hyperfine interactions
\begin{equation}
    \label{eq: isotropic contact hyperfine}
    \mathcal{H}_\mathrm{n}=a_\mathrm{iso}(\mathbf{I}\cdot \mathbf{S})-\mu_\mathrm{n} g_\mathrm{I} (\mathbf{I}\cdot \mathbf{B}) \,.
\end{equation} 
We determine that an isotropic contact hyperfine parameter of $a_\mathrm{iso}=$\qty{2.75(0.03)}{\micro eV} is sufficient to fit the data, as shown in \cref{fig: hyperfine spectrum}. Since only the levels within the $\mathcal{J}=1$ manifold are observed, the cost function across the 54-dimensional manifold is minimized through an unbalanced linear sum assignment method \cite{Crouse2016OnAlgorithms}. $g_\mathcal{L}$, $g_\mathrm{s}$, and $\lambda$ are fixed from our high-field measurements. The nuclear magneton $\mu_\mathrm{n}$ and g-factor are fixed with $g_\mathrm{I}=1.456$\,\cite{chemLin}. The magnetic orbital, dipole, and electric quadrupole interactions are significantly smaller than the contact term and may be quenched similar to the orbital Zeeman term \cite{Abragam1951TheoryCrystals,Abragam_and_Bleaney}. Within the two-electron manifold, the electron that remains unpromoted to the 1s:T$_\mathrm{2}$ configuration will contribute to the contact hyperfine with at most half the total spin density much like the 2$^3$p two electron configuration of helium-like ions in free space. The nuclear Zeeman interaction, though present in our calculations, is negligible.  The complete parameters describing the LS-coupled manifold are summarized in \cref{table: Excited state params}. The separation of $\mathrm{m}_\mathcal{J}=-1$ into 6 distinct states shows a high degree of agreement with our fit, confirming that we have optically resolved the hyperfine structure in this metastable excited state. 

\begin{table}[b]
    \centering
    \renewcommand*\arraystretch{1.3}
    \centering 
    \begin{tabular}{c c c c}  
        \hline \hline
        Parameter                &Description                         &Value        &Units\\
        \hline
        $E_\mathrm{s}$           &Unperturbed singlet energy          &796.14(1)    &meV  \\ 
        $E_\mathrm{t}$           &Unperturbed triplet energy          &774.91(1)    &meV  \\ 
        $\Delta_\mathrm{st}$     &Exchange splitting                  &21.23(1)     &meV  \\
        $\lambda$                &Effective spin-orbit strength       &47.6(9)      &\qty{}{\micro eV} \\
        $a_\mathrm{iso}$         &Contact hyperfine strength          &2.75(3)      &\qty{}{\micro eV} \\
        $g_\mathrm{s}$           &Spin Land\'e g-factor               &2            &- \\
        $g_{\mathcal{L}}$        &Effective orbital g-factor          &0            &- \\
        \hline \hline
    \end{tabular} 
    \caption{1s:$\mathrm{A_1}$\,1s:$\mathrm{T_2}$ excited state Hamiltonian parameters. The Land\'e $g$-factors are assumed parameters, and were not fit to the data.}
    \label{table: Excited state params}
\end{table}

\section{Discussion}
Electron spins in the metastable excited states of point defects have been proposed as ancillary qubits for nuclear spin initialization and readout in quantum repeater nodes \cite{Lee2013ReadoutAncilla,Deak2024QuantumBit}. Unlike \Aliplus{}, previously proposed silicon colour centres with diamagnetic ground states, including C \cite{Wen2025OpticalSpin}, G \cite{Watkins1982}, and C$\mathrm{_i}$ \cite{Deak2024QuantumBit}, do not exhibit an LS-coupled excited state. Like \Aliplus{}, they have triplet and singlet excited states; however, their reduced symmetry splits the triplet into the magnetic quantum numbers $m_\mathrm{s} \in \{0, \pm1\}$. A number of encoding schemes have been proposed for these levels such as using the $m_\mathrm{s}=\pm1$ states as the computational states $\ket{0}$ and $\ket{1}$ and performing geometric phase gates through the ancillary $\mathrm{m_s=0}$ state \cite{Ivanov2022EffectLocalization}. Spin initialization is performed by pumping the nuclear-spin-preserving singlet transition allowing for the respective hyperfine levels to populate within the triplet. Coherent control of an equivalent system in diamond has been demonstrated including spin polarization transfer from the electron to the nuclear spin through an avoided crossing \cite{Lee2013ReadoutAncilla}. In this demonstration, since each $\mathrm{m_s}$ level decays at a different rate, the ODMR contrast results in direct measurement of the electron state. 

\Aliplus{} offers an alternative approach. The spin-orbit manifold forms a ladder system with the $\mathcal{J}=0$ or $2$ levels serving as possible shelving states. The relative splitting to $\mathcal{J}=1$, given by $\lambda$ and $2\lambda$, is accessible with microwave electronics. The nuclear Zeeman-split states in the ground state and within $\mathcal{J}=0$ may be driven by radio-frequency magnetic fields.  

Furthermore, decays from $\mathcal{J}=0$ to the ground state are forbidden by symmetry. Decays from $\mathcal{J}=2$ to the ground state are weakly allowed through qudrupole-like interactions. The lifetime of the $\mathcal{J}=1$ bright state is, therefore, much shorter than these shelving states, but these decay pathways relax as the manifold forms an admixture of effective angular momentum states in an intermediate magnetic field regime suitable for spin control. 

Optical readout of the nuclear spin is enabled by exciting \singlet{}, which preferentially decays to \triplet{}$_{(\mathcal{J}=1)}$. Spin-resolved emission back to the ground state implements a projective measurement of the nuclear spin. We calculate the eigenstates of the excited state Hamiltonian, including both the electron Zeeman and spin-orbit from \cref{eq: electron Zeeman Hamiltonian} and the contact hyperfine and nuclear Zeeman terms from \cref{eq: isotropic contact hyperfine}, in the uncoupled Zeeman spin basis. We label the hyperfine states of \triplet{} at field by the dominant spin components $\ket{\mathcal{J}, m_\mathcal{J}, m_\mathrm{I}}$ and note that these eigenstates are mixed in the uncoupled Zeeman basis. The decay probability from each of the $\mathcal{J}=1$ levels to a given nuclear Zeeman level in the ground state $\ket{m'_\mathrm{I}}$ is approximated as the nuclear spin overlap, or partial inner product,
\begin{equation}
P({m_\mathcal{J}}, m_\mathrm{I};m'_\mathrm{I})\approx\braket{\mathcal{J}=1, m_\mathcal{J}, m_\mathrm{I}| m'_\mathrm{I}} \,.
\end{equation}
The nuclear branching ratio  
\begin{equation}
    F({m_\mathcal{J}}, m_\mathrm{I};m'_\mathrm{I})=\frac{P(m_\mathcal{J}, m_\mathrm{I};m'_\mathrm{I})}{\sum_k P(m_\mathcal{J}, m_\mathrm{I};k)}
    \label{eq: nuclear branching ratio}
\end{equation}
is normalized by all decay pathways. We find that $\norm{\mathbf{B}} \geq \qty{606}{\text{mT}}$ is sufficient to achieve $>99$\% nuclear-spin-preserving transitions, suitable for single-shot readout. At this field, the average spectral separation of the nuclear spin-selective decays from the $m_\mathcal{J}=-1,0$ and 1 sub-branches are \qty{539.0}{MHz}, \qty{206.8}{MHz}, and \qty{116.8}{MHz} respectively. These are resolvable by, for example, fibre Fabry-Perot cavities which are commercially available at this wavelength with bandwidths of \qty{50}{MHz}. Nuclear spin initialization may be performed by measurement and feedback to the nuclear spin via spin control in the ground state. Alternatively, the hyperfine splitting in the \triplet{} state may allow for spin control of the nuclear spin during its long lifetime. 

In principle, the emission of the triplet state may be applied in optical remote entanglement protocols \cite{Barrett2005}. However, a short excited state lifetime is desirable for quantum networking, and the lifetime of the metastable \triplet{} state ($\tau_{\mathrm{t}} =\qty{25.5(0.5)}{ms}$) limits the feasible readout, initialization, and entanglement rates. Although an optical nanocavity can enhance a selected transition, the rates would remain slow even at an effective Purcell factor of 1,000 \cite{Merkel2020CoherentPurcell}. Since the SOC relaxes the spin selection rule between the triplet and ground states, this transition will be faster in heavier ions as shown by the large increase in absorption strength of this transition with increasing atomic weight for the neutral chalcogens in silicon\cite{Peale1988ZeemanSilicon}. Strain tuning may allow further mixing of the states \cite{Bergman1988}, and the heavier group-III ions, indium and thallium, are believed to exist in interstitial lattice positions \cite{Baron1969ElectricalSilicon}. If these heavier group-III acceptors can be maintained in interstitial sites, the singly ionized states of the resulting donors may provide faster optical emission better-suited for optical spin readout and networking. 

Throughout these studies, we observed additional transitions in absorption of aluminum-related defects and the thermal transformation of \Aliplus{} into the previously observed AM2 series \cite{Latushko1989IRSilicon}. Further results on the AM2 He$^{+}$-like donor series, the B1 and B2 absorption transitions reported by Latuschko \textit{et al}. \cite{Latushko1989IRSilicon, Latushko1989IdentificationSi:Al}, and new \Ali-related neutral and doubly ionized donor absorption series will be detailed in a forthcoming publication.

\section{Conclusion}
Our results provide a comprehensive characterization of the \Aliplus{} centre in $\mathrm{^{28}Si}$, observing emission from both the \stateT{} excited state triplet and singlet in the telecom L and C bands, respectively. In this isotopically purified host, the transitions are narrow enough to resolve both the fine and hyperfine structure of the triplet state \triplet{}. These measurements are the first optically-resolved hyperfine structure within the excited state of a silicon colour centre with a diamagnetic ground state. We predict that these transitions are >99\% nuclear spin-preserving at feasible magnetic fields and that multi-electron donors similar to \Aliplus{} may be used as spin-photon interfaces, capable of optical initialization and readout of the nuclear spin.  

\begin{acknowledgements}
This work was supported by the Canadian Department of National Defence (DND) through the innovation for Defence Excellence and Security (IDEaS), the Natural Sciences and Engineering Research Council of Canada (NSERC), the New Frontiers in Research Fund (NFRF), the  Canada research chair program (CRC), the Canada Foundation of Innovation (CFI), the B.C. Knowledge Development Fund (BCKDF), and the Quantum Information Science program at the Canadian Institute for Advanced Research (CIFAR).
\newline

\textit{Data Availability} --- The data that supports the findings of this article is available from the authors upon reasonable request.
\end{acknowledgements}


\begin{thebibliography}{63}%
\makeatletter
\providecommand \@ifxundefined [1]{%
 \@ifx{#1\undefined}
}%
\providecommand \@ifnum [1]{%
 \ifnum #1\expandafter \@firstoftwo
 \else \expandafter \@secondoftwo
 \fi
}%
\providecommand \@ifx [1]{%
 \ifx #1\expandafter \@firstoftwo
 \else \expandafter \@secondoftwo
 \fi
}%
\providecommand \natexlab [1]{#1}%
\providecommand \enquote  [1]{``#1''}%
\providecommand \bibnamefont  [1]{#1}%
\providecommand \bibfnamefont [1]{#1}%
\providecommand \citenamefont [1]{#1}%
\providecommand \href@noop [0]{\@secondoftwo}%
\providecommand \href [0]{\begingroup \@sanitize@url \@href}%
\providecommand \@href[1]{\@@startlink{#1}\@@href}%
\providecommand \@@href[1]{\endgroup#1\@@endlink}%
\providecommand \@sanitize@url [0]{\catcode `\\12\catcode `\$12\catcode `\&12\catcode `\#12\catcode `\^12\catcode `\_12\catcode `\%12\relax}%
\providecommand \@@startlink[1]{}%
\providecommand \@@endlink[0]{}%
\providecommand \url  [0]{\begingroup\@sanitize@url \@url }%
\providecommand \@url [1]{\endgroup\@href {#1}{\urlprefix }}%
\providecommand \urlprefix  [0]{URL }%
\providecommand \Eprint [0]{\href }%
\providecommand \doibase [0]{http://dx.doi.org/}%
\providecommand \selectlanguage [0]{\@gobble}%
\providecommand \bibinfo  [0]{\@secondoftwo}%
\providecommand \bibfield  [0]{\@secondoftwo}%
\providecommand \translation [1]{[#1]}%
\providecommand \BibitemOpen [0]{}%
\providecommand \bibitemStop [0]{}%
\providecommand \bibitemNoStop [0]{.\EOS\space}%
\providecommand \EOS [0]{\spacefactor3000\relax}%
\providecommand \BibitemShut  [1]{\csname bibitem#1\endcsname}%
\let\auto@bib@innerbib\@empty
\bibitem [{\citenamefont {Tyryshkin}\ \emph {et~al.}(2012)\citenamefont {Tyryshkin}, \citenamefont {Tojo}, \citenamefont {Morton}, \citenamefont {Riemann}, \citenamefont {Abrosimov}, \citenamefont {Becker}, \citenamefont {Pohl}, \citenamefont {Schenkel}, \citenamefont {Thewalt}, \citenamefont {Itoh},\ and\ \citenamefont {Lyon}}]{Tyryshkin2012}%
  \BibitemOpen
  \bibfield  {author} {\bibinfo {author} {\bibfnamefont {A.~M.}\ \bibnamefont {Tyryshkin}}, \bibinfo {author} {\bibfnamefont {S.}~\bibnamefont {Tojo}}, \bibinfo {author} {\bibfnamefont {J.~J.}\ \bibnamefont {Morton}}, \bibinfo {author} {\bibfnamefont {H.}~\bibnamefont {Riemann}}, \bibinfo {author} {\bibfnamefont {N.~V.}\ \bibnamefont {Abrosimov}}, \bibinfo {author} {\bibfnamefont {P.}~\bibnamefont {Becker}}, \bibinfo {author} {\bibfnamefont {H.~J.}\ \bibnamefont {Pohl}}, \bibinfo {author} {\bibfnamefont {T.}~\bibnamefont {Schenkel}}, \bibinfo {author} {\bibfnamefont {M.~L.}\ \bibnamefont {Thewalt}}, \bibinfo {author} {\bibfnamefont {K.~M.}\ \bibnamefont {Itoh}}, \ and\ \bibinfo {author} {\bibfnamefont {S.~A.}\ \bibnamefont {Lyon}},\ }\href {\doibase 10.1038/nmat3182} {\bibfield  {journal} {\bibinfo  {journal} {Nature Materials}\ }\textbf {\bibinfo {volume} {11}},\ \bibinfo {pages} {143} (\bibinfo {year} {2012})}\BibitemShut {NoStop}%
\bibitem [{\citenamefont {Steger}\ \emph {et~al.}(2012)\citenamefont {Steger}, \citenamefont {Saeedi}, \citenamefont {Thewalt}, \citenamefont {Morton}, \citenamefont {Riemann}, \citenamefont {Abrosimov}, \citenamefont {Becker},\ and\ \citenamefont {Pohl}}]{Steger2012a}%
  \BibitemOpen
  \bibfield  {author} {\bibinfo {author} {\bibfnamefont {M.}~\bibnamefont {Steger}}, \bibinfo {author} {\bibfnamefont {K.}~\bibnamefont {Saeedi}}, \bibinfo {author} {\bibfnamefont {M.~L.}\ \bibnamefont {Thewalt}}, \bibinfo {author} {\bibfnamefont {J.~J.}\ \bibnamefont {Morton}}, \bibinfo {author} {\bibfnamefont {H.}~\bibnamefont {Riemann}}, \bibinfo {author} {\bibfnamefont {N.~V.}\ \bibnamefont {Abrosimov}}, \bibinfo {author} {\bibfnamefont {P.}~\bibnamefont {Becker}}, \ and\ \bibinfo {author} {\bibfnamefont {H.~J.}\ \bibnamefont {Pohl}},\ }\href {\doibase 10.1126/SCIENCE.1217635} {\bibfield  {journal} {\bibinfo  {journal} {Science}\ }\textbf {\bibinfo {volume} {336}},\ \bibinfo {pages} {1280} (\bibinfo {year} {2012})}\BibitemShut {NoStop}%
\bibitem [{\citenamefont {Saeedi}\ \emph {et~al.}(2013)\citenamefont {Saeedi}, \citenamefont {Simmons}, \citenamefont {Salvail}, \citenamefont {Dluhy}, \citenamefont {Riemann}, \citenamefont {Abromosimov}, \citenamefont {Becker}, \citenamefont {Pohl}, \citenamefont {Morton},\ and\ \citenamefont {THewalt}}]{Saeedi2013}%
  \BibitemOpen
  \bibfield  {author} {\bibinfo {author} {\bibfnamefont {K.}~\bibnamefont {Saeedi}}, \bibinfo {author} {\bibfnamefont {S.}~\bibnamefont {Simmons}}, \bibinfo {author} {\bibfnamefont {J.~Z.}\ \bibnamefont {Salvail}}, \bibinfo {author} {\bibfnamefont {P.}~\bibnamefont {Dluhy}}, \bibinfo {author} {\bibfnamefont {H.}~\bibnamefont {Riemann}}, \bibinfo {author} {\bibfnamefont {N.~V.}\ \bibnamefont {Abromosimov}}, \bibinfo {author} {\bibfnamefont {P.}~\bibnamefont {Becker}}, \bibinfo {author} {\bibfnamefont {H.-J.}\ \bibnamefont {Pohl}}, \bibinfo {author} {\bibfnamefont {J.~J.~L.}\ \bibnamefont {Morton}}, \ and\ \bibinfo {author} {\bibfnamefont {M.~L.~W.}\ \bibnamefont {THewalt}},\ }\href {\doibase 10.1126/science.1239584} {\bibfield  {journal} {\bibinfo  {journal} {Science}\ }\textbf {\bibinfo {volume} {342}},\ \bibinfo {pages} {830} (\bibinfo {year} {2013})}\BibitemShut {NoStop}%
\bibitem [{\citenamefont {Buckley}\ \emph {et~al.}(2017)\citenamefont {Buckley}, \citenamefont {Chiles}, \citenamefont {McCaughan}, \citenamefont {Moody}, \citenamefont {Silverman}, \citenamefont {Stevens}, \citenamefont {Mirin}, \citenamefont {Nam},\ and\ \citenamefont {Shainline}}]{Buckley2017}%
  \BibitemOpen
  \bibfield  {author} {\bibinfo {author} {\bibfnamefont {S.}~\bibnamefont {Buckley}}, \bibinfo {author} {\bibfnamefont {J.}~\bibnamefont {Chiles}}, \bibinfo {author} {\bibfnamefont {A.~N.}\ \bibnamefont {McCaughan}}, \bibinfo {author} {\bibfnamefont {G.}~\bibnamefont {Moody}}, \bibinfo {author} {\bibfnamefont {K.~L.}\ \bibnamefont {Silverman}}, \bibinfo {author} {\bibfnamefont {M.~J.}\ \bibnamefont {Stevens}}, \bibinfo {author} {\bibfnamefont {R.~P.}\ \bibnamefont {Mirin}}, \bibinfo {author} {\bibfnamefont {S.~W.}\ \bibnamefont {Nam}}, \ and\ \bibinfo {author} {\bibfnamefont {J.~M.}\ \bibnamefont {Shainline}},\ }\href {\doibase 10.1063/1.4994692} {\bibfield  {journal} {\bibinfo  {journal} {Applied Physics Letters}\ }\textbf {\bibinfo {volume} {111}},\ \bibinfo {pages} {141101} (\bibinfo {year} {2017})}\BibitemShut {NoStop}%
\bibitem [{\citenamefont {Gritsch}\ \emph {et~al.}(2023)\citenamefont {Gritsch}, \citenamefont {Ulanowski},\ and\ \citenamefont {Reiserer}}]{Gritsch2023Purcell}%
  \BibitemOpen
  \bibfield  {author} {\bibinfo {author} {\bibfnamefont {A.}~\bibnamefont {Gritsch}}, \bibinfo {author} {\bibfnamefont {A.}~\bibnamefont {Ulanowski}}, \ and\ \bibinfo {author} {\bibfnamefont {A.}~\bibnamefont {Reiserer}},\ }\href {\doibase 10.1364/optica.486167} {\bibfield  {journal} {\bibinfo  {journal} {Optica}\ }\textbf {\bibinfo {volume} {10}},\ \bibinfo {pages} {783} (\bibinfo {year} {2023})}\BibitemShut {NoStop}%
\bibitem [{\citenamefont {Johnston}\ \emph {et~al.}(2024)\citenamefont {Johnston}, \citenamefont {Felix-Rendon}, \citenamefont {Wong},\ and\ \citenamefont {Chen}}]{Johnston_2024_CavityCoupledTcentre}%
  \BibitemOpen
  \bibfield  {author} {\bibinfo {author} {\bibfnamefont {A.}~\bibnamefont {Johnston}}, \bibinfo {author} {\bibfnamefont {U.}~\bibnamefont {Felix-Rendon}}, \bibinfo {author} {\bibfnamefont {Y.-E.}\ \bibnamefont {Wong}}, \ and\ \bibinfo {author} {\bibfnamefont {S.}~\bibnamefont {Chen}},\ }\href {\doibase 10.1038/s41467-024-46643-8} {\bibfield  {journal} {\bibinfo  {journal} {Nature Communications}\ }\textbf {\bibinfo {volume} {15}},\ \bibinfo {pages} {2350} (\bibinfo {year} {2024})}\BibitemShut {NoStop}%
\bibitem [{\citenamefont {Lefaucher}\ \emph {et~al.}(2025)\citenamefont {Lefaucher}, \citenamefont {Baron}, \citenamefont {Jager}, \citenamefont {Calvo}, \citenamefont {Els{\"{a}}sser}, \citenamefont {Coppola}, \citenamefont {Mazen}, \citenamefont {Kerdil{\`{e}}s}, \citenamefont {Cache}, \citenamefont {Dr{\'{e}}au},\ and\ \citenamefont {G{\'{e}}rard}}]{Lefaucher2025BrightMicrocavity}%
  \BibitemOpen
  \bibfield  {author} {\bibinfo {author} {\bibfnamefont {B.}~\bibnamefont {Lefaucher}}, \bibinfo {author} {\bibfnamefont {Y.}~\bibnamefont {Baron}}, \bibinfo {author} {\bibfnamefont {J.-B.}\ \bibnamefont {Jager}}, \bibinfo {author} {\bibfnamefont {V.}~\bibnamefont {Calvo}}, \bibinfo {author} {\bibfnamefont {C.}~\bibnamefont {Els{\"{a}}sser}}, \bibinfo {author} {\bibfnamefont {G.}~\bibnamefont {Coppola}}, \bibinfo {author} {\bibfnamefont {F.}~\bibnamefont {Mazen}}, \bibinfo {author} {\bibfnamefont {S.}~\bibnamefont {Kerdil{\`{e}}s}}, \bibinfo {author} {\bibfnamefont {F.}~\bibnamefont {Cache}}, \bibinfo {author} {\bibfnamefont {A.}~\bibnamefont {Dr{\'{e}}au}}, \ and\ \bibinfo {author} {\bibfnamefont {J.-M.}\ \bibnamefont {G{\'{e}}rard}},\ }\href {http://arxiv.org/abs/2501.12744} {\bibfield  {journal} {\bibinfo  {journal} {arXiv:2501.12744}\ } (\bibinfo {year} {2025})}\BibitemShut {NoStop}%
\bibitem [{\citenamefont {Komza}\ \emph {et~al.}(2024)\citenamefont {Komza}, \citenamefont {Samutpraphoot}, \citenamefont {Odeh}, \citenamefont {Tang}, \citenamefont {Mathew}, \citenamefont {Chang}, \citenamefont {Song}, \citenamefont {Kim}, \citenamefont {Xiong}, \citenamefont {Hautier},\ and\ \citenamefont {Sipahigil}}]{Komza2024IndistinguishablePhotonics}%
  \BibitemOpen
  \bibfield  {author} {\bibinfo {author} {\bibfnamefont {L.}~\bibnamefont {Komza}}, \bibinfo {author} {\bibfnamefont {P.}~\bibnamefont {Samutpraphoot}}, \bibinfo {author} {\bibfnamefont {M.}~\bibnamefont {Odeh}}, \bibinfo {author} {\bibfnamefont {Y.~L.}\ \bibnamefont {Tang}}, \bibinfo {author} {\bibfnamefont {M.}~\bibnamefont {Mathew}}, \bibinfo {author} {\bibfnamefont {J.}~\bibnamefont {Chang}}, \bibinfo {author} {\bibfnamefont {H.}~\bibnamefont {Song}}, \bibinfo {author} {\bibfnamefont {M.~K.}\ \bibnamefont {Kim}}, \bibinfo {author} {\bibfnamefont {Y.}~\bibnamefont {Xiong}}, \bibinfo {author} {\bibfnamefont {G.}~\bibnamefont {Hautier}}, \ and\ \bibinfo {author} {\bibfnamefont {A.}~\bibnamefont {Sipahigil}},\ }\href {\doibase 10.1038/s41467-024-51265-1} {\bibfield  {journal} {\bibinfo  {journal} {Nature Communications 2024 15:1}\ }\textbf {\bibinfo {volume} {15}},\ \bibinfo {pages} {1} (\bibinfo {year} {2024})}\BibitemShut {NoStop}%
\bibitem [{\citenamefont {Afzal}\ \emph {et~al.}(2024)\citenamefont {Afzal}, \citenamefont {Akhlaghi}, \citenamefont {Beale}, \citenamefont {Bedroya}, \citenamefont {Bell}, \citenamefont {Bergeron}, \citenamefont {Bonsma-Fisher}, \citenamefont {Bychkova}, \citenamefont {Chaisson}, \citenamefont {Chartrand}, \citenamefont {Clear}, \citenamefont {Darcie}, \citenamefont {DeAbreu}, \citenamefont {DeLisle}, \citenamefont {Duncan}, \citenamefont {Smith}, \citenamefont {Dunn}, \citenamefont {Ebrahimi}, \citenamefont {Evetts}, \citenamefont {Pinheiro}, \citenamefont {Fuentes}, \citenamefont {Georgiou}, \citenamefont {Guha}, \citenamefont {Haenel}, \citenamefont {Higginbottom}, \citenamefont {Jackson}, \citenamefont {Jahed}, \citenamefont {Khorshidahmad}, \citenamefont {Shandilya}, \citenamefont {Kurkjian}, \citenamefont {Lauk}, \citenamefont {Lee-Hone}, \citenamefont {Lin}, \citenamefont {Litynskyy}, \citenamefont {Lock}, \citenamefont {Ma}, \citenamefont {MacGilp}, \citenamefont {MacQuarrie}, \citenamefont {Mar},
  \citenamefont {Khah}, \citenamefont {Matiash}, \citenamefont {Meyer-Scott}, \citenamefont {Michaels}, \citenamefont {Motira}, \citenamefont {Noori}, \citenamefont {Ospadov}, \citenamefont {Patel}, \citenamefont {Patscheider}, \citenamefont {Paulson}, \citenamefont {Petruk}, \citenamefont {Ravindranath}, \citenamefont {Reznychenko}, \citenamefont {Ruether}, \citenamefont {Ruscica}, \citenamefont {Saxena}, \citenamefont {Schaller}, \citenamefont {Seidlitz}, \citenamefont {Senger}, \citenamefont {Lee}, \citenamefont {Sevoyan}, \citenamefont {Simmons}, \citenamefont {Soykal}, \citenamefont {Stott}, \citenamefont {Tran}, \citenamefont {Tserkis}, \citenamefont {Ulhaq}, \citenamefont {Vine}, \citenamefont {Weeks}, \citenamefont {Wolfowicz},\ and\ \citenamefont {Yoneda}}]{Afzal2024}%
  \BibitemOpen
  \bibfield  {author} {\bibinfo {author} {\bibfnamefont {F.}~\bibnamefont {Afzal}}, \bibinfo {author} {\bibfnamefont {M.}~\bibnamefont {Akhlaghi}}, \bibinfo {author} {\bibfnamefont {S.~J.}\ \bibnamefont {Beale}}, \bibinfo {author} {\bibfnamefont {O.}~\bibnamefont {Bedroya}}, \bibinfo {author} {\bibfnamefont {K.}~\bibnamefont {Bell}}, \bibinfo {author} {\bibfnamefont {L.}~\bibnamefont {Bergeron}}, \bibinfo {author} {\bibfnamefont {K.}~\bibnamefont {Bonsma-Fisher}}, \bibinfo {author} {\bibfnamefont {P.}~\bibnamefont {Bychkova}}, \bibinfo {author} {\bibfnamefont {Z.~M.~E.}\ \bibnamefont {Chaisson}}, \bibinfo {author} {\bibfnamefont {C.}~\bibnamefont {Chartrand}}, \bibinfo {author} {\bibfnamefont {C.}~\bibnamefont {Clear}}, \bibinfo {author} {\bibfnamefont {A.}~\bibnamefont {Darcie}}, \bibinfo {author} {\bibfnamefont {A.}~\bibnamefont {DeAbreu}}, \bibinfo {author} {\bibfnamefont {C.}~\bibnamefont {DeLisle}}, \bibinfo {author} {\bibfnamefont {L.~A.}\ \bibnamefont {Duncan}}, \bibinfo {author} {\bibfnamefont
  {C.~D.}\ \bibnamefont {Smith}}, \bibinfo {author} {\bibfnamefont {J.}~\bibnamefont {Dunn}}, \bibinfo {author} {\bibfnamefont {A.}~\bibnamefont {Ebrahimi}}, \bibinfo {author} {\bibfnamefont {N.}~\bibnamefont {Evetts}}, \bibinfo {author} {\bibfnamefont {D.~F.}\ \bibnamefont {Pinheiro}}, \bibinfo {author} {\bibfnamefont {P.}~\bibnamefont {Fuentes}}, \bibinfo {author} {\bibfnamefont {T.}~\bibnamefont {Georgiou}}, \bibinfo {author} {\bibfnamefont {B.}~\bibnamefont {Guha}}, \bibinfo {author} {\bibfnamefont {R.}~\bibnamefont {Haenel}}, \bibinfo {author} {\bibfnamefont {D.}~\bibnamefont {Higginbottom}}, \bibinfo {author} {\bibfnamefont {D.~M.}\ \bibnamefont {Jackson}}, \bibinfo {author} {\bibfnamefont {N.}~\bibnamefont {Jahed}}, \bibinfo {author} {\bibfnamefont {A.}~\bibnamefont {Khorshidahmad}}, \bibinfo {author} {\bibfnamefont {P.~K.}\ \bibnamefont {Shandilya}}, \bibinfo {author} {\bibfnamefont {A.~T.~K.}\ \bibnamefont {Kurkjian}}, \bibinfo {author} {\bibfnamefont {N.}~\bibnamefont {Lauk}}, \bibinfo {author}
  {\bibfnamefont {N.~R.}\ \bibnamefont {Lee-Hone}}, \bibinfo {author} {\bibfnamefont {E.}~\bibnamefont {Lin}}, \bibinfo {author} {\bibfnamefont {R.}~\bibnamefont {Litynskyy}}, \bibinfo {author} {\bibfnamefont {D.}~\bibnamefont {Lock}}, \bibinfo {author} {\bibfnamefont {L.}~\bibnamefont {Ma}}, \bibinfo {author} {\bibfnamefont {I.}~\bibnamefont {MacGilp}}, \bibinfo {author} {\bibfnamefont {E.~R.}\ \bibnamefont {MacQuarrie}}, \bibinfo {author} {\bibfnamefont {A.}~\bibnamefont {Mar}}, \bibinfo {author} {\bibfnamefont {A.~M.}\ \bibnamefont {Khah}}, \bibinfo {author} {\bibfnamefont {A.}~\bibnamefont {Matiash}}, \bibinfo {author} {\bibfnamefont {E.}~\bibnamefont {Meyer-Scott}}, \bibinfo {author} {\bibfnamefont {C.~P.}\ \bibnamefont {Michaels}}, \bibinfo {author} {\bibfnamefont {J.}~\bibnamefont {Motira}}, \bibinfo {author} {\bibfnamefont {N.~K.}\ \bibnamefont {Noori}}, \bibinfo {author} {\bibfnamefont {E.}~\bibnamefont {Ospadov}}, \bibinfo {author} {\bibfnamefont {E.}~\bibnamefont {Patel}}, \bibinfo {author}
  {\bibfnamefont {A.}~\bibnamefont {Patscheider}}, \bibinfo {author} {\bibfnamefont {D.}~\bibnamefont {Paulson}}, \bibinfo {author} {\bibfnamefont {A.}~\bibnamefont {Petruk}}, \bibinfo {author} {\bibfnamefont {A.~L.}\ \bibnamefont {Ravindranath}}, \bibinfo {author} {\bibfnamefont {B.}~\bibnamefont {Reznychenko}}, \bibinfo {author} {\bibfnamefont {M.}~\bibnamefont {Ruether}}, \bibinfo {author} {\bibfnamefont {J.}~\bibnamefont {Ruscica}}, \bibinfo {author} {\bibfnamefont {K.}~\bibnamefont {Saxena}}, \bibinfo {author} {\bibfnamefont {Z.}~\bibnamefont {Schaller}}, \bibinfo {author} {\bibfnamefont {A.}~\bibnamefont {Seidlitz}}, \bibinfo {author} {\bibfnamefont {J.}~\bibnamefont {Senger}}, \bibinfo {author} {\bibfnamefont {Y.~S.}\ \bibnamefont {Lee}}, \bibinfo {author} {\bibfnamefont {O.}~\bibnamefont {Sevoyan}}, \bibinfo {author} {\bibfnamefont {S.}~\bibnamefont {Simmons}}, \bibinfo {author} {\bibfnamefont {O.}~\bibnamefont {Soykal}}, \bibinfo {author} {\bibfnamefont {L.}~\bibnamefont {Stott}}, \bibinfo {author}
  {\bibfnamefont {Q.}~\bibnamefont {Tran}}, \bibinfo {author} {\bibfnamefont {S.}~\bibnamefont {Tserkis}}, \bibinfo {author} {\bibfnamefont {A.}~\bibnamefont {Ulhaq}}, \bibinfo {author} {\bibfnamefont {W.}~\bibnamefont {Vine}}, \bibinfo {author} {\bibfnamefont {R.}~\bibnamefont {Weeks}}, \bibinfo {author} {\bibfnamefont {G.}~\bibnamefont {Wolfowicz}}, \ and\ \bibinfo {author} {\bibfnamefont {I.}~\bibnamefont {Yoneda}},\ }\href {http://arxiv.org/abs/2406.01704} {\bibfield  {journal} {\bibinfo  {journal} {arXiv 2406.01704}\ } (\bibinfo {year} {2024})}\BibitemShut {NoStop}%
\bibitem [{\citenamefont {Dobinson}\ \emph {et~al.}(2025)\citenamefont {Dobinson}, \citenamefont {Bowness}, \citenamefont {Meynell}, \citenamefont {Chartrand}, \citenamefont {Hoffmann}, \citenamefont {Gascoine}, \citenamefont {MacGilp}, \citenamefont {Afzal}, \citenamefont {Dangel}, \citenamefont {Jahed}, \citenamefont {Thewalt}, \citenamefont {Simmons},\ and\ \citenamefont {Higginbottom}}]{dobinson2025Electrically}%
  \BibitemOpen
  \bibfield  {author} {\bibinfo {author} {\bibfnamefont {M.}~\bibnamefont {Dobinson}}, \bibinfo {author} {\bibfnamefont {C.}~\bibnamefont {Bowness}}, \bibinfo {author} {\bibfnamefont {S.~A.}\ \bibnamefont {Meynell}}, \bibinfo {author} {\bibfnamefont {C.}~\bibnamefont {Chartrand}}, \bibinfo {author} {\bibfnamefont {E.}~\bibnamefont {Hoffmann}}, \bibinfo {author} {\bibfnamefont {M.}~\bibnamefont {Gascoine}}, \bibinfo {author} {\bibfnamefont {I.}~\bibnamefont {MacGilp}}, \bibinfo {author} {\bibfnamefont {F.}~\bibnamefont {Afzal}}, \bibinfo {author} {\bibfnamefont {C.}~\bibnamefont {Dangel}}, \bibinfo {author} {\bibfnamefont {N.}~\bibnamefont {Jahed}}, \bibinfo {author} {\bibfnamefont {M.~L.~W.}\ \bibnamefont {Thewalt}}, \bibinfo {author} {\bibfnamefont {S.}~\bibnamefont {Simmons}}, \ and\ \bibinfo {author} {\bibfnamefont {D.~B.}\ \bibnamefont {Higginbottom}},\ }\href {\doibase https://doi.org/10.1038/s41566-025-01752-8} {\bibfield  {journal} {\bibinfo  {journal} {Nature Photonics}\ }\textbf {\bibinfo {volume}
  {19}},\ \bibinfo {pages} {1132} (\bibinfo {year} {2025})}\BibitemShut {NoStop}%
\bibitem [{\citenamefont {Day}\ \emph {et~al.}(2025)\citenamefont {Day}, \citenamefont {Zhang}, \citenamefont {Jin}, \citenamefont {Song}, \citenamefont {Sutula}, \citenamefont {Sipahigil}, \citenamefont {Bhaskar},\ and\ \citenamefont {Hu}}]{day2025Probing}%
  \BibitemOpen
  \bibfield  {author} {\bibinfo {author} {\bibfnamefont {A.~M.}\ \bibnamefont {Day}}, \bibinfo {author} {\bibfnamefont {C.}~\bibnamefont {Zhang}}, \bibinfo {author} {\bibfnamefont {C.}~\bibnamefont {Jin}}, \bibinfo {author} {\bibfnamefont {H.}~\bibnamefont {Song}}, \bibinfo {author} {\bibfnamefont {M.}~\bibnamefont {Sutula}}, \bibinfo {author} {\bibfnamefont {A.}~\bibnamefont {Sipahigil}}, \bibinfo {author} {\bibfnamefont {M.~K.}\ \bibnamefont {Bhaskar}}, \ and\ \bibinfo {author} {\bibfnamefont {E.~L.}\ \bibnamefont {Hu}},\ }\href {http://arxiv.org/abs/2501.11888} {\bibfield  {journal} {\bibinfo  {journal} {arXiv:2501.11888}\ } (\bibinfo {year} {2025})}\BibitemShut {NoStop}%
\bibitem [{\citenamefont {Song}\ \emph {et~al.}(2025)\citenamefont {Song}, \citenamefont {Zhang}, \citenamefont {Komza}, \citenamefont {Fiaschi}, \citenamefont {Xiong}, \citenamefont {Zhi}, \citenamefont {Dhuey}, \citenamefont {Schwartzberg}, \citenamefont {Schenkel}, \citenamefont {Hautier}, \citenamefont {Zhang},\ and\ \citenamefont {Sipahigil}}]{Song2025LongLived}%
  \BibitemOpen
  \bibfield  {author} {\bibinfo {author} {\bibfnamefont {H.}~\bibnamefont {Song}}, \bibinfo {author} {\bibfnamefont {X.}~\bibnamefont {Zhang}}, \bibinfo {author} {\bibfnamefont {L.}~\bibnamefont {Komza}}, \bibinfo {author} {\bibfnamefont {N.}~\bibnamefont {Fiaschi}}, \bibinfo {author} {\bibfnamefont {Y.}~\bibnamefont {Xiong}}, \bibinfo {author} {\bibfnamefont {Y.}~\bibnamefont {Zhi}}, \bibinfo {author} {\bibfnamefont {S.}~\bibnamefont {Dhuey}}, \bibinfo {author} {\bibfnamefont {A.}~\bibnamefont {Schwartzberg}}, \bibinfo {author} {\bibfnamefont {T.}~\bibnamefont {Schenkel}}, \bibinfo {author} {\bibfnamefont {G.}~\bibnamefont {Hautier}}, \bibinfo {author} {\bibfnamefont {Z.-H.}\ \bibnamefont {Zhang}}, \ and\ \bibinfo {author} {\bibfnamefont {A.}~\bibnamefont {Sipahigil}},\ }\href {http://arxiv.org/abs/2504.15467} {\bibfield  {journal} {\bibinfo  {journal} {arXiv:2504.15467}\ } (\bibinfo {year} {2025})}\BibitemShut {NoStop}%
\bibitem [{\citenamefont {Benjamin}\ \emph {et~al.}(2006)\citenamefont {Benjamin}, \citenamefont {Browne}, \citenamefont {Fitzsimons},\ and\ \citenamefont {Morton}}]{Benjamin2006BrokeredComputation}%
  \BibitemOpen
  \bibfield  {author} {\bibinfo {author} {\bibfnamefont {S.~C.}\ \bibnamefont {Benjamin}}, \bibinfo {author} {\bibfnamefont {D.~E.}\ \bibnamefont {Browne}}, \bibinfo {author} {\bibfnamefont {J.}~\bibnamefont {Fitzsimons}}, \ and\ \bibinfo {author} {\bibfnamefont {J.~J.}\ \bibnamefont {Morton}},\ }\href {\doibase 10.1088/1367-2630/8/8/141} {\bibfield  {journal} {\bibinfo  {journal} {New Journal of Physics}\ }\textbf {\bibinfo {volume} {8}},\ \bibinfo {pages} {0} (\bibinfo {year} {2006})}\BibitemShut {NoStop}%
\bibitem [{\citenamefont {Pompili}\ \emph {et~al.}(2021)\citenamefont {Pompili}, \citenamefont {Hermans}, \citenamefont {Baier}, \citenamefont {Beukers}, \citenamefont {Humphreys}, \citenamefont {Schouten}, \citenamefont {Vermeulen}, \citenamefont {Tiggelman}, \citenamefont {dos Santos~Martins}, \citenamefont {Dirkse}, \citenamefont {Wehner},\ and\ \citenamefont {Hanson}}]{Pompili2021a}%
  \BibitemOpen
  \bibfield  {author} {\bibinfo {author} {\bibfnamefont {M.}~\bibnamefont {Pompili}}, \bibinfo {author} {\bibfnamefont {S.~L.}\ \bibnamefont {Hermans}}, \bibinfo {author} {\bibfnamefont {S.}~\bibnamefont {Baier}}, \bibinfo {author} {\bibfnamefont {H.~K.}\ \bibnamefont {Beukers}}, \bibinfo {author} {\bibfnamefont {P.~C.}\ \bibnamefont {Humphreys}}, \bibinfo {author} {\bibfnamefont {R.~N.}\ \bibnamefont {Schouten}}, \bibinfo {author} {\bibfnamefont {R.~F.}\ \bibnamefont {Vermeulen}}, \bibinfo {author} {\bibfnamefont {M.~J.}\ \bibnamefont {Tiggelman}}, \bibinfo {author} {\bibfnamefont {L.}~\bibnamefont {dos Santos~Martins}}, \bibinfo {author} {\bibfnamefont {B.}~\bibnamefont {Dirkse}}, \bibinfo {author} {\bibfnamefont {S.}~\bibnamefont {Wehner}}, \ and\ \bibinfo {author} {\bibfnamefont {R.}~\bibnamefont {Hanson}},\ }\href {\doibase 10.1126/science.abg1919} {\bibfield  {journal} {\bibinfo  {journal} {Science}\ }\textbf {\bibinfo {volume} {372}},\ \bibinfo {pages} {259} (\bibinfo {year} {2021})}\BibitemShut
  {NoStop}%
\bibitem [{\citenamefont {Kane}(1998)}]{Kane1998}%
  \BibitemOpen
  \bibfield  {author} {\bibinfo {author} {\bibfnamefont {B.~E.}\ \bibnamefont {Kane}},\ }\href {\doibase 10.1038/30156} {\bibfield  {journal} {\bibinfo  {journal} {Nature}\ }\textbf {\bibinfo {volume} {393}},\ \bibinfo {pages} {133} (\bibinfo {year} {1998})}\BibitemShut {NoStop}%
\bibitem [{\citenamefont {Yang}\ \emph {et~al.}(2022)\citenamefont {Yang}, \citenamefont {Fan}, \citenamefont {Zhang}, \citenamefont {Duan}, \citenamefont {De~Boo}, \citenamefont {Ahlefeldt}, \citenamefont {Longdell}, \citenamefont {Johnson}, \citenamefont {McCallum}, \citenamefont {Sellars}, \citenamefont {Rogge}, \citenamefont {Yin},\ and\ \citenamefont {Du}}]{Yang2022}%
  \BibitemOpen
  \bibfield  {author} {\bibinfo {author} {\bibfnamefont {J.}~\bibnamefont {Yang}}, \bibinfo {author} {\bibfnamefont {W.}~\bibnamefont {Fan}}, \bibinfo {author} {\bibfnamefont {Y.}~\bibnamefont {Zhang}}, \bibinfo {author} {\bibfnamefont {C.}~\bibnamefont {Duan}}, \bibinfo {author} {\bibfnamefont {G.~G.}\ \bibnamefont {De~Boo}}, \bibinfo {author} {\bibfnamefont {R.~L.}\ \bibnamefont {Ahlefeldt}}, \bibinfo {author} {\bibfnamefont {J.~J.}\ \bibnamefont {Longdell}}, \bibinfo {author} {\bibfnamefont {B.~C.}\ \bibnamefont {Johnson}}, \bibinfo {author} {\bibfnamefont {J.~C.}\ \bibnamefont {McCallum}}, \bibinfo {author} {\bibfnamefont {M.~J.}\ \bibnamefont {Sellars}}, \bibinfo {author} {\bibfnamefont {S.}~\bibnamefont {Rogge}}, \bibinfo {author} {\bibfnamefont {C.}~\bibnamefont {Yin}}, \ and\ \bibinfo {author} {\bibfnamefont {J.}~\bibnamefont {Du}},\ }\href {\doibase 10.1103/PhysRevB.105.235306} {\bibfield  {journal} {\bibinfo  {journal} {Phys. Rev. B}\ }\textbf {\bibinfo {volume} {105}},\ \bibinfo {pages} {235306}
  (\bibinfo {year} {2022})}\BibitemShut {NoStop}%
\bibitem [{\citenamefont {Deabreu}\ \emph {et~al.}(2019)\citenamefont {Deabreu}, \citenamefont {Bowness}, \citenamefont {Abraham}, \citenamefont {Medvedova}, \citenamefont {Morse}, \citenamefont {Riemann}, \citenamefont {Abrosimov}, \citenamefont {Becker}, \citenamefont {Pohl}, \citenamefont {Thewalt},\ and\ \citenamefont {Simmons}}]{Deabreu2019}%
  \BibitemOpen
  \bibfield  {author} {\bibinfo {author} {\bibfnamefont {A.}~\bibnamefont {Deabreu}}, \bibinfo {author} {\bibfnamefont {C.}~\bibnamefont {Bowness}}, \bibinfo {author} {\bibfnamefont {R.~J.}\ \bibnamefont {Abraham}}, \bibinfo {author} {\bibfnamefont {A.}~\bibnamefont {Medvedova}}, \bibinfo {author} {\bibfnamefont {K.~J.}\ \bibnamefont {Morse}}, \bibinfo {author} {\bibfnamefont {H.}~\bibnamefont {Riemann}}, \bibinfo {author} {\bibfnamefont {N.~V.}\ \bibnamefont {Abrosimov}}, \bibinfo {author} {\bibfnamefont {P.}~\bibnamefont {Becker}}, \bibinfo {author} {\bibfnamefont {H.~J.}\ \bibnamefont {Pohl}}, \bibinfo {author} {\bibfnamefont {M.~L.}\ \bibnamefont {Thewalt}}, \ and\ \bibinfo {author} {\bibfnamefont {S.}~\bibnamefont {Simmons}},\ }\href {\doibase 10.1103/PhysRevApplied.11.044036} {\bibfield  {journal} {\bibinfo  {journal} {Phys. Rev. Appl.}\ }\textbf {\bibinfo {volume} {11}},\ \bibinfo {pages} {44036} (\bibinfo {year} {2019})}\BibitemShut {NoStop}%
\bibitem [{\citenamefont {Bergeron}\ \emph {et~al.}(2020)\citenamefont {Bergeron}, \citenamefont {Chartrand}, \citenamefont {Kurkjian}, \citenamefont {Morse}, \citenamefont {Riemann}, \citenamefont {Abrosimov}, \citenamefont {Becker}, \citenamefont {Pohl}, \citenamefont {Thewalt},\ and\ \citenamefont {Simmons}}]{Bergeron:2020_PRX}%
  \BibitemOpen
  \bibfield  {author} {\bibinfo {author} {\bibfnamefont {L.}~\bibnamefont {Bergeron}}, \bibinfo {author} {\bibfnamefont {C.}~\bibnamefont {Chartrand}}, \bibinfo {author} {\bibfnamefont {A.~T.~K.}\ \bibnamefont {Kurkjian}}, \bibinfo {author} {\bibfnamefont {K.~J.}\ \bibnamefont {Morse}}, \bibinfo {author} {\bibfnamefont {H.}~\bibnamefont {Riemann}}, \bibinfo {author} {\bibfnamefont {N.~V.}\ \bibnamefont {Abrosimov}}, \bibinfo {author} {\bibfnamefont {P.}~\bibnamefont {Becker}}, \bibinfo {author} {\bibfnamefont {H.-J.}\ \bibnamefont {Pohl}}, \bibinfo {author} {\bibfnamefont {M.~L.~W.}\ \bibnamefont {Thewalt}}, \ and\ \bibinfo {author} {\bibfnamefont {S.}~\bibnamefont {Simmons}},\ }\href {\doibase 10.1103/prxquantum.1.020301} {\bibfield  {journal} {\bibinfo  {journal} {PRX Quantum}\ }\textbf {\bibinfo {volume} {1}},\ \bibinfo {pages} {20301} (\bibinfo {year} {2020})}\BibitemShut {NoStop}%
\bibitem [{\citenamefont {Clear}\ \emph {et~al.}(2024)\citenamefont {Clear}, \citenamefont {Hosseini}, \citenamefont {AlizadehKhaledi}, \citenamefont {Brunelle}, \citenamefont {Woolverton}, \citenamefont {Kanaganayagam}, \citenamefont {Kazemi}, \citenamefont {Chartrand}, \citenamefont {Keshavarz}, \citenamefont {Xiong}, \citenamefont {Alaerts}, \citenamefont {Soykal}, \citenamefont {Hautier}, \citenamefont {Karassiouk}, \citenamefont {Thewalt}, \citenamefont {Higginbottom},\ and\ \citenamefont {Simmons}}]{Clear2024optical}%
  \BibitemOpen
  \bibfield  {author} {\bibinfo {author} {\bibfnamefont {C.}~\bibnamefont {Clear}}, \bibinfo {author} {\bibfnamefont {S.}~\bibnamefont {Hosseini}}, \bibinfo {author} {\bibfnamefont {A.}~\bibnamefont {AlizadehKhaledi}}, \bibinfo {author} {\bibfnamefont {N.}~\bibnamefont {Brunelle}}, \bibinfo {author} {\bibfnamefont {A.}~\bibnamefont {Woolverton}}, \bibinfo {author} {\bibfnamefont {J.}~\bibnamefont {Kanaganayagam}}, \bibinfo {author} {\bibfnamefont {M.}~\bibnamefont {Kazemi}}, \bibinfo {author} {\bibfnamefont {C.}~\bibnamefont {Chartrand}}, \bibinfo {author} {\bibfnamefont {M.}~\bibnamefont {Keshavarz}}, \bibinfo {author} {\bibfnamefont {Y.}~\bibnamefont {Xiong}}, \bibinfo {author} {\bibfnamefont {L.}~\bibnamefont {Alaerts}}, \bibinfo {author} {\bibfnamefont {{\"O}.~O.}\ \bibnamefont {Soykal}}, \bibinfo {author} {\bibfnamefont {G.}~\bibnamefont {Hautier}}, \bibinfo {author} {\bibfnamefont {V.}~\bibnamefont {Karassiouk}}, \bibinfo {author} {\bibfnamefont {M.}~\bibnamefont {Thewalt}}, \bibinfo {author}
  {\bibfnamefont {D.}~\bibnamefont {Higginbottom}}, \ and\ \bibinfo {author} {\bibfnamefont {S.}~\bibnamefont {Simmons}},\ }\href {\doibase 10.1103/PhysRevApplied.22.064014} {\bibfield  {journal} {\bibinfo  {journal} {Physical Review Applied}\ }\textbf {\bibinfo {volume} {22}},\ \bibinfo {pages} {064014} (\bibinfo {year} {2024})}\BibitemShut {NoStop}%
\bibitem [{\citenamefont {O'Donnell}\ \emph {et~al.}(1982)\citenamefont {O'Donnell}, \citenamefont {Lee}, \citenamefont {Watkins},\ and\ \citenamefont {{K. P. O'Donnell}}}]{ODonnell1982}%
  \BibitemOpen
  \bibfield  {author} {\bibinfo {author} {\bibfnamefont {K.}~\bibnamefont {O'Donnell}}, \bibinfo {author} {\bibfnamefont {K.}~\bibnamefont {Lee}}, \bibinfo {author} {\bibfnamefont {G.}~\bibnamefont {Watkins}}, \ and\ \bibinfo {author} {\bibnamefont {{K. P. O'Donnell}}},\ }\href {\doibase 10.1016/0378-4363(83)90256-5} {\bibfield  {journal} {\bibinfo  {journal} {Physica B+C}\ }\textbf {\bibinfo {volume} {116}},\ \bibinfo {pages} {258} (\bibinfo {year} {1982})}\BibitemShut {NoStop}%
\bibitem [{\citenamefont {Lee}\ \emph {et~al.}(1982)\citenamefont {Lee}, \citenamefont {O'Donnell}, \citenamefont {Weber}, \citenamefont {Cavenett},\ and\ \citenamefont {Watkins}}]{Watkins1982}%
  \BibitemOpen
  \bibfield  {author} {\bibinfo {author} {\bibfnamefont {K.~M.}\ \bibnamefont {Lee}}, \bibinfo {author} {\bibfnamefont {K.~P.}\ \bibnamefont {O'Donnell}}, \bibinfo {author} {\bibfnamefont {J.}~\bibnamefont {Weber}}, \bibinfo {author} {\bibfnamefont {B.~C.}\ \bibnamefont {Cavenett}}, \ and\ \bibinfo {author} {\bibfnamefont {G.~D.}\ \bibnamefont {Watkins}},\ }\href {\doibase 10.1103/physrevlett.48.37} {\bibfield  {journal} {\bibinfo  {journal} {Phys. Rev. Lett.}\ }\textbf {\bibinfo {volume} {48}},\ \bibinfo {pages} {37} (\bibinfo {year} {1982})}\BibitemShut {NoStop}%
\bibitem [{\citenamefont {Beaufils}\ \emph {et~al.}(2018)\citenamefont {Beaufils}, \citenamefont {Redjem}, \citenamefont {Rousseau}, \citenamefont {Jacques}, \citenamefont {Kuznetsov}, \citenamefont {Raynaud}, \citenamefont {Voisin}, \citenamefont {Benali}, \citenamefont {Herzig}, \citenamefont {Pezzagna}, \citenamefont {Meijer}, \citenamefont {Abbarchi},\ and\ \citenamefont {Cassabois}}]{Beaufils2018}%
  \BibitemOpen
  \bibfield  {author} {\bibinfo {author} {\bibfnamefont {C.}~\bibnamefont {Beaufils}}, \bibinfo {author} {\bibfnamefont {W.}~\bibnamefont {Redjem}}, \bibinfo {author} {\bibfnamefont {E.}~\bibnamefont {Rousseau}}, \bibinfo {author} {\bibfnamefont {V.}~\bibnamefont {Jacques}}, \bibinfo {author} {\bibfnamefont {A.~Y.}\ \bibnamefont {Kuznetsov}}, \bibinfo {author} {\bibfnamefont {C.}~\bibnamefont {Raynaud}}, \bibinfo {author} {\bibfnamefont {C.}~\bibnamefont {Voisin}}, \bibinfo {author} {\bibfnamefont {A.}~\bibnamefont {Benali}}, \bibinfo {author} {\bibfnamefont {T.}~\bibnamefont {Herzig}}, \bibinfo {author} {\bibfnamefont {S.}~\bibnamefont {Pezzagna}}, \bibinfo {author} {\bibfnamefont {J.}~\bibnamefont {Meijer}}, \bibinfo {author} {\bibfnamefont {M.}~\bibnamefont {Abbarchi}}, \ and\ \bibinfo {author} {\bibfnamefont {G.}~\bibnamefont {Cassabois}},\ }\href {\doibase 10.1103/PhysRevB.97.035303} {\bibfield  {journal} {\bibinfo  {journal} {Physical Review B}\ }\textbf {\bibinfo {volume} {97}},\ \bibinfo {pages}
  {035303} (\bibinfo {year} {2018})}\BibitemShut {NoStop}%
\bibitem [{\citenamefont {Cache}\ \emph {et~al.}(2025)\citenamefont {Cache}, \citenamefont {Baron}, \citenamefont {Lefaucher}, \citenamefont {Jager}, \citenamefont {Mazen}, \citenamefont {Mil{\'{e}}si}, \citenamefont {Kerdil{\`{e}}s}, \citenamefont {Robert-Philip}, \citenamefont {G{\'{e}}rard}, \citenamefont {Cassabois}, \citenamefont {Jacques},\ and\ \citenamefont {Dr{\'{e}}au}}]{Cache2025}%
  \BibitemOpen
  \bibfield  {author} {\bibinfo {author} {\bibfnamefont {F.}~\bibnamefont {Cache}}, \bibinfo {author} {\bibfnamefont {Y.}~\bibnamefont {Baron}}, \bibinfo {author} {\bibfnamefont {B.}~\bibnamefont {Lefaucher}}, \bibinfo {author} {\bibfnamefont {J.-B.}\ \bibnamefont {Jager}}, \bibinfo {author} {\bibfnamefont {F.}~\bibnamefont {Mazen}}, \bibinfo {author} {\bibfnamefont {F.}~\bibnamefont {Mil{\'{e}}si}}, \bibinfo {author} {\bibfnamefont {S.}~\bibnamefont {Kerdil{\`{e}}s}}, \bibinfo {author} {\bibfnamefont {I.}~\bibnamefont {Robert-Philip}}, \bibinfo {author} {\bibfnamefont {J.-M.}\ \bibnamefont {G{\'{e}}rard}}, \bibinfo {author} {\bibfnamefont {G.}~\bibnamefont {Cassabois}}, \bibinfo {author} {\bibfnamefont {V.}~\bibnamefont {Jacques}}, \ and\ \bibinfo {author} {\bibfnamefont {A.}~\bibnamefont {Dr{\'{e}}au}},\ }\href {http://arxiv.org/abs/2510.15590} {\bibfield  {journal} {\bibinfo  {journal} {arXiv:2510.15590}\ } (\bibinfo {year} {2025})}\BibitemShut {NoStop}%
\bibitem [{\citenamefont {Cache}\ \emph {et~al.}(2026)\citenamefont {Cache}, \citenamefont {R.}, \citenamefont {Herzig}, \citenamefont {Kuznetsov}, \citenamefont {Pezzagna}, \citenamefont {Abbarchi}, \citenamefont {Robert-Philip}, \citenamefont {G{\'{e}}rard}, \citenamefont {Cassabois}, \citenamefont {Jacques},\ and\ \citenamefont {Dr{\'{e}}au}}]{Cache2026OpticalSilicon}%
  \BibitemOpen
  \bibfield  {author} {\bibinfo {author} {\bibfnamefont {F.}~\bibnamefont {Cache}}, \bibinfo {author} {\bibfnamefont {K.~V.}\ \bibnamefont {R.}}, \bibinfo {author} {\bibfnamefont {T.}~\bibnamefont {Herzig}}, \bibinfo {author} {\bibfnamefont {A.~Y.}\ \bibnamefont {Kuznetsov}}, \bibinfo {author} {\bibfnamefont {S.}~\bibnamefont {Pezzagna}}, \bibinfo {author} {\bibfnamefont {M.}~\bibnamefont {Abbarchi}}, \bibinfo {author} {\bibfnamefont {I.}~\bibnamefont {Robert-Philip}}, \bibinfo {author} {\bibfnamefont {J.-M.}\ \bibnamefont {G{\'{e}}rard}}, \bibinfo {author} {\bibfnamefont {G.}~\bibnamefont {Cassabois}}, \bibinfo {author} {\bibfnamefont {V.}~\bibnamefont {Jacques}}, \ and\ \bibinfo {author} {\bibfnamefont {A.}~\bibnamefont {Dr{\'{e}}au}},\ }\href {http://arxiv.org/abs/2605.12473} {\  (\bibinfo {year} {2026})}\BibitemShut {NoStop}%
\bibitem [{\citenamefont {Udvarhelyi}\ \emph {et~al.}(2021)\citenamefont {Udvarhelyi}, \citenamefont {Somogyi}, \citenamefont {Thiering},\ and\ \citenamefont {Gali}}]{Udvarhelyi2021}%
  \BibitemOpen
  \bibfield  {author} {\bibinfo {author} {\bibfnamefont {P.}~\bibnamefont {Udvarhelyi}}, \bibinfo {author} {\bibfnamefont {B.}~\bibnamefont {Somogyi}}, \bibinfo {author} {\bibfnamefont {G.}~\bibnamefont {Thiering}}, \ and\ \bibinfo {author} {\bibfnamefont {A.}~\bibnamefont {Gali}},\ }\href {\doibase https://doi.org/10.1103/PhysRevLett.127.196402} {\bibfield  {journal} {\bibinfo  {journal} {Phys. Rev. Lett.}\ }\textbf {\bibinfo {volume} {127}},\ \bibinfo {pages} {196402} (\bibinfo {year} {2021})}\BibitemShut {NoStop}%
\bibitem [{\citenamefont {Ishikawa}\ \emph {et~al.}(2011)\citenamefont {Ishikawa}, \citenamefont {Koga}, \citenamefont {Itahashi}, \citenamefont {Itoh},\ and\ \citenamefont {Vlasenko}}]{Ishikawa2011OpticalSilicon}%
  \BibitemOpen
  \bibfield  {author} {\bibinfo {author} {\bibfnamefont {T.}~\bibnamefont {Ishikawa}}, \bibinfo {author} {\bibfnamefont {K.}~\bibnamefont {Koga}}, \bibinfo {author} {\bibfnamefont {T.}~\bibnamefont {Itahashi}}, \bibinfo {author} {\bibfnamefont {K.~M.}\ \bibnamefont {Itoh}}, \ and\ \bibinfo {author} {\bibfnamefont {L.~S.}\ \bibnamefont {Vlasenko}},\ }\href {\doibase 10.1103/PhysRevB.84.115204} {\bibfield  {journal} {\bibinfo  {journal} {Phys. Rev. B}\ }\textbf {\bibinfo {volume} {84}} (\bibinfo {year} {2011}),\ 10.1103/PhysRevB.84.115204}\BibitemShut {NoStop}%
\bibitem [{\citenamefont {Wen}\ \emph {et~al.}(2025)\citenamefont {Wen}, \citenamefont {Pieplow}, \citenamefont {Yang}, \citenamefont {Jamshidi}, \citenamefont {Helm}, \citenamefont {Luo}, \citenamefont {Schr{\"{o}}der}, \citenamefont {Zhou},\ and\ \citenamefont {Berenc{\'{e}}n}}]{Wen2025OpticalSpin}%
  \BibitemOpen
  \bibfield  {author} {\bibinfo {author} {\bibfnamefont {S.}~\bibnamefont {Wen}}, \bibinfo {author} {\bibfnamefont {G.}~\bibnamefont {Pieplow}}, \bibinfo {author} {\bibfnamefont {J.}~\bibnamefont {Yang}}, \bibinfo {author} {\bibfnamefont {K.}~\bibnamefont {Jamshidi}}, \bibinfo {author} {\bibfnamefont {M.}~\bibnamefont {Helm}}, \bibinfo {author} {\bibfnamefont {J.-W.}\ \bibnamefont {Luo}}, \bibinfo {author} {\bibfnamefont {T.}~\bibnamefont {Schr{\"{o}}der}}, \bibinfo {author} {\bibfnamefont {S.}~\bibnamefont {Zhou}}, \ and\ \bibinfo {author} {\bibfnamefont {Y.}~\bibnamefont {Berenc{\'{e}}n}},\ }\href {https://arxiv.org/abs/2502.07632} {\bibfield  {journal} {\bibinfo  {journal} {arXiv:2502.07632}\ } (\bibinfo {year} {2025})}\BibitemShut {NoStop}%
\bibitem [{\citenamefont {De{\'{a}}k}\ \emph {et~al.}(2024)\citenamefont {De{\'{a}}k}, \citenamefont {Li},\ and\ \citenamefont {Gali}}]{Deak2024QuantumBit}%
  \BibitemOpen
  \bibfield  {author} {\bibinfo {author} {\bibfnamefont {P.}~\bibnamefont {De{\'{a}}k}}, \bibinfo {author} {\bibfnamefont {S.}~\bibnamefont {Li}}, \ and\ \bibinfo {author} {\bibfnamefont {A.}~\bibnamefont {Gali}},\ }\href {\doibase 10.1038/s42005-024-01834-z} {\bibfield  {journal} {\bibinfo  {journal} {Communications Physics}\ }\textbf {\bibinfo {volume} {7}} (\bibinfo {year} {2024}),\ 10.1038/s42005-024-01834-z}\BibitemShut {NoStop}%
\bibitem [{\citenamefont {Jhuria}\ \emph {et~al.}(2024)\citenamefont {Jhuria}, \citenamefont {Ivanov}, \citenamefont {Polley}, \citenamefont {Zhiyenbayev}, \citenamefont {Liu}, \citenamefont {Persaud}, \citenamefont {Redjem}, \citenamefont {Qarony}, \citenamefont {Parajuli}, \citenamefont {Ji}, \citenamefont {Gonsalves}, \citenamefont {Bokor}, \citenamefont {Tan}, \citenamefont {Kant{\'{e}}},\ and\ \citenamefont {Schenkel}}]{Jhuria2024ProgrammableQuantum}%
  \BibitemOpen
  \bibfield  {author} {\bibinfo {author} {\bibfnamefont {K.}~\bibnamefont {Jhuria}}, \bibinfo {author} {\bibfnamefont {V.}~\bibnamefont {Ivanov}}, \bibinfo {author} {\bibfnamefont {D.}~\bibnamefont {Polley}}, \bibinfo {author} {\bibfnamefont {Y.}~\bibnamefont {Zhiyenbayev}}, \bibinfo {author} {\bibfnamefont {W.}~\bibnamefont {Liu}}, \bibinfo {author} {\bibfnamefont {A.}~\bibnamefont {Persaud}}, \bibinfo {author} {\bibfnamefont {W.}~\bibnamefont {Redjem}}, \bibinfo {author} {\bibfnamefont {W.}~\bibnamefont {Qarony}}, \bibinfo {author} {\bibfnamefont {P.}~\bibnamefont {Parajuli}}, \bibinfo {author} {\bibfnamefont {Q.}~\bibnamefont {Ji}}, \bibinfo {author} {\bibfnamefont {A.~J.}\ \bibnamefont {Gonsalves}}, \bibinfo {author} {\bibfnamefont {J.}~\bibnamefont {Bokor}}, \bibinfo {author} {\bibfnamefont {L.~Z.}\ \bibnamefont {Tan}}, \bibinfo {author} {\bibfnamefont {B.}~\bibnamefont {Kant{\'{e}}}}, \ and\ \bibinfo {author} {\bibfnamefont {T.}~\bibnamefont {Schenkel}},\ }\href {\doibase 10.1038/s41467-024-48714-2}
  {\bibfield  {journal} {\bibinfo  {journal} {Nature Communications 2024 15:1}\ }\textbf {\bibinfo {volume} {15}},\ \bibinfo {pages} {1} (\bibinfo {year} {2024})}\BibitemShut {NoStop}%
\bibitem [{\citenamefont {Lee}\ \emph {et~al.}(2013)\citenamefont {Lee}, \citenamefont {Widmann}, \citenamefont {Rendler}, \citenamefont {Doherty}, \citenamefont {Babinec}, \citenamefont {Yang}, \citenamefont {Eyer}, \citenamefont {Siyushev}, \citenamefont {Hausmann}, \citenamefont {Loncar}, \citenamefont {Bodrog}, \citenamefont {Gali}, \citenamefont {Manson}, \citenamefont {Fedder},\ and\ \citenamefont {Wrachtrup}}]{Lee2013ReadoutAncilla}%
  \BibitemOpen
  \bibfield  {author} {\bibinfo {author} {\bibfnamefont {S.~Y.}\ \bibnamefont {Lee}}, \bibinfo {author} {\bibfnamefont {M.}~\bibnamefont {Widmann}}, \bibinfo {author} {\bibfnamefont {T.}~\bibnamefont {Rendler}}, \bibinfo {author} {\bibfnamefont {M.~W.}\ \bibnamefont {Doherty}}, \bibinfo {author} {\bibfnamefont {T.~M.}\ \bibnamefont {Babinec}}, \bibinfo {author} {\bibfnamefont {S.}~\bibnamefont {Yang}}, \bibinfo {author} {\bibfnamefont {M.}~\bibnamefont {Eyer}}, \bibinfo {author} {\bibfnamefont {P.}~\bibnamefont {Siyushev}}, \bibinfo {author} {\bibfnamefont {B.~J.}\ \bibnamefont {Hausmann}}, \bibinfo {author} {\bibfnamefont {M.}~\bibnamefont {Loncar}}, \bibinfo {author} {\bibfnamefont {Z.}~\bibnamefont {Bodrog}}, \bibinfo {author} {\bibfnamefont {A.}~\bibnamefont {Gali}}, \bibinfo {author} {\bibfnamefont {N.~B.}\ \bibnamefont {Manson}}, \bibinfo {author} {\bibfnamefont {H.}~\bibnamefont {Fedder}}, \ and\ \bibinfo {author} {\bibfnamefont {J.}~\bibnamefont {Wrachtrup}},\ }\href {\doibase 10.1038/nnano.2013.104}
  {\bibfield  {journal} {\bibinfo  {journal} {Nature Nanotechnology}\ }\textbf {\bibinfo {volume} {8}},\ \bibinfo {pages} {487} (\bibinfo {year} {2013})}\BibitemShut {NoStop}%
\bibitem [{\citenamefont {Akhtar}\ \emph {et~al.}(2012)\citenamefont {Akhtar}, \citenamefont {Filidou}, \citenamefont {Sekiguchi}, \citenamefont {Kawakami}, \citenamefont {Itahashi}, \citenamefont {Vlasenko}, \citenamefont {Morton},\ and\ \citenamefont {Itoh}}]{Akhtar2012CoherentSilicon}%
  \BibitemOpen
  \bibfield  {author} {\bibinfo {author} {\bibfnamefont {W.}~\bibnamefont {Akhtar}}, \bibinfo {author} {\bibfnamefont {V.}~\bibnamefont {Filidou}}, \bibinfo {author} {\bibfnamefont {T.}~\bibnamefont {Sekiguchi}}, \bibinfo {author} {\bibfnamefont {E.}~\bibnamefont {Kawakami}}, \bibinfo {author} {\bibfnamefont {T.}~\bibnamefont {Itahashi}}, \bibinfo {author} {\bibfnamefont {L.}~\bibnamefont {Vlasenko}}, \bibinfo {author} {\bibfnamefont {J.~J.~L.}\ \bibnamefont {Morton}}, \ and\ \bibinfo {author} {\bibfnamefont {K.~M.}\ \bibnamefont {Itoh}},\ }\href {\doibase 10.1103/PhysRevLett.108.097601} {\bibfield  {journal} {\bibinfo  {journal} {Phys. Rev. Lett.}\ }\textbf {\bibinfo {volume} {108}},\ \bibinfo {pages} {097601} (\bibinfo {year} {2012})}\BibitemShut {NoStop}%
\bibitem [{\citenamefont {Latushko}\ \emph {et~al.}(1982)\citenamefont {Latushko}, \citenamefont {Nassur},\ and\ \citenamefont {Petrov}}]{Latushko1982IRAluminum}%
  \BibitemOpen
  \bibfield  {author} {\bibinfo {author} {\bibfnamefont {Y.~I.}\ \bibnamefont {Latushko}}, \bibinfo {author} {\bibfnamefont {F.}~\bibnamefont {Nassur}}, \ and\ \bibinfo {author} {\bibfnamefont {V.~V.}\ \bibnamefont {Petrov}},\ }\href@noop {} {\emph {\bibinfo {title} {Zhurnal Prikladnoi Spektroskopii}}},\ \bibinfo {type} {Tech. Rep.}\ \bibinfo {number} {0}\ (\bibinfo {year} {1982})\BibitemShut {NoStop}%
\bibitem [{\citenamefont {Latushko}\ and\ \citenamefont {Petrov}(1989{\natexlab{a}})}]{Latushko1989IRSilicon}%
  \BibitemOpen
  \bibfield  {author} {\bibinfo {author} {\bibfnamefont {Y.}~\bibnamefont {Latushko}}\ and\ \bibinfo {author} {\bibfnamefont {V.}~\bibnamefont {Petrov}},\ }\href {\doibase 10.4028/www.scientific.net/MSF.38-41.1169} {\bibfield  {journal} {\bibinfo  {journal} {Materials Science Forum}\ }\textbf {\bibinfo {volume} {38-41}},\ \bibinfo {pages} {1169} (\bibinfo {year} {1989}{\natexlab{a}})}\BibitemShut {NoStop}%
\bibitem [{\citenamefont {Latushko}\ and\ \citenamefont {Petrov}(1989{\natexlab{b}})}]{Latushko1989IdentificationSi:Al}%
  \BibitemOpen
  \bibfield  {author} {\bibinfo {author} {\bibfnamefont {Y.}~\bibnamefont {Latushko}}\ and\ \bibinfo {author} {\bibfnamefont {V.}~\bibnamefont {Petrov}},\ }\href {\doibase 10.1557/PROC-163-277} {\bibfield  {journal} {\bibinfo  {journal} {MRS Proceedings}\ }\textbf {\bibinfo {volume} {163}},\ \bibinfo {pages} {277} (\bibinfo {year} {1989}{\natexlab{b}})}\BibitemShut {NoStop}%
\bibitem [{\citenamefont {Grimmeiss}\ \emph {et~al.}(1982)\citenamefont {Grimmeiss}, \citenamefont {Janz{\'{e}}n},\ and\ \citenamefont {Larsson}}]{Grimmeiss1982}%
  \BibitemOpen
  \bibfield  {author} {\bibinfo {author} {\bibfnamefont {H.~G.}\ \bibnamefont {Grimmeiss}}, \bibinfo {author} {\bibfnamefont {E.}~\bibnamefont {Janz{\'{e}}n}}, \ and\ \bibinfo {author} {\bibfnamefont {K.}~\bibnamefont {Larsson}},\ }\href {\doibase 10.1103/PhysRevB.25.2627} {\bibfield  {journal} {\bibinfo  {journal} {Phys. Rev. B}\ }\textbf {\bibinfo {volume} {25}},\ \bibinfo {pages} {2627} (\bibinfo {year} {1982})}\BibitemShut {NoStop}%
\bibitem [{\citenamefont {Bergman}\ \emph {et~al.}(1988)\citenamefont {Bergman}, \citenamefont {Grossmann}, \citenamefont {Grimmeiss}, \citenamefont {Stavola}, \citenamefont {Holm},\ and\ \citenamefont {Wagner}}]{Bergman1988}%
  \BibitemOpen
  \bibfield  {author} {\bibinfo {author} {\bibfnamefont {K.}~\bibnamefont {Bergman}}, \bibinfo {author} {\bibfnamefont {G.}~\bibnamefont {Grossmann}}, \bibinfo {author} {\bibfnamefont {H.}~\bibnamefont {Grimmeiss}}, \bibinfo {author} {\bibfnamefont {M.}~\bibnamefont {Stavola}}, \bibinfo {author} {\bibfnamefont {C.}~\bibnamefont {Holm}}, \ and\ \bibinfo {author} {\bibfnamefont {P.}~\bibnamefont {Wagner}},\ }\href {\doibase https://doi.org/10.1103/PhysRevB.37.10738} {\bibfield  {journal} {\bibinfo  {journal} {Phys. Rev. B.}\ }\textbf {\bibinfo {volume} {37}},\ \bibinfo {pages} {10738} (\bibinfo {year} {1988})}\BibitemShut {NoStop}%
\bibitem [{\citenamefont {Bergman}\ \emph {et~al.}(1986)\citenamefont {Bergman}, \citenamefont {Grossmann}, \citenamefont {Grimmeiss},\ and\ \citenamefont {Stavola}}]{Bergman1986ObservationSilicon}%
  \BibitemOpen
  \bibfield  {author} {\bibinfo {author} {\bibfnamefont {K.}~\bibnamefont {Bergman}}, \bibinfo {author} {\bibfnamefont {G.}~\bibnamefont {Grossmann}}, \bibinfo {author} {\bibfnamefont {H.~G.}\ \bibnamefont {Grimmeiss}}, \ and\ \bibinfo {author} {\bibfnamefont {M.}~\bibnamefont {Stavola}},\ }\href {\doibase 10.1103/PhysRevLett.56.2827} {\bibfield  {journal} {\bibinfo  {journal} {Phys. Rev. Lett.}\ }\textbf {\bibinfo {volume} {56}} (\bibinfo {year} {1986}),\ 10.1103/PhysRevLett.56.2827}\BibitemShut {NoStop}%
\bibitem [{\citenamefont {Janzen}\ \emph {et~al.}(1984)\citenamefont {Janzen}, \citenamefont {Stedman}, \citenamefont {Grossmann},\ and\ \citenamefont {Grimmeiss}}]{Janzen1984}%
  \BibitemOpen
  \bibfield  {author} {\bibinfo {author} {\bibfnamefont {E.}~\bibnamefont {Janzen}}, \bibinfo {author} {\bibfnamefont {R.}~\bibnamefont {Stedman}}, \bibinfo {author} {\bibfnamefont {G.}~\bibnamefont {Grossmann}}, \ and\ \bibinfo {author} {\bibfnamefont {H.}~\bibnamefont {Grimmeiss}},\ }\href {\doibase https://doi.org/10.1103/PhysRevB.29.1907} {\bibfield  {journal} {\bibinfo  {journal} {Phys. Rev. B.}\ }\textbf {\bibinfo {volume} {29}},\ \bibinfo {pages} {1907} (\bibinfo {year} {1984})}\BibitemShut {NoStop}%
\bibitem [{\citenamefont {Steger}\ \emph {et~al.}(2009)\citenamefont {Steger}, \citenamefont {Yang}, \citenamefont {Thewalt}, \citenamefont {Cardona}, \citenamefont {Riemann}, \citenamefont {Abrosimov}, \citenamefont {Churbanov}, \citenamefont {Gusev}, \citenamefont {Bulanov}, \citenamefont {Kovalev}, \citenamefont {Kaliteevskii}, \citenamefont {Godisov}, \citenamefont {Becker}, \citenamefont {Pohl}, \citenamefont {Haller},\ and\ \citenamefont {Ager}}]{Steger2009}%
  \BibitemOpen
  \bibfield  {author} {\bibinfo {author} {\bibfnamefont {M.}~\bibnamefont {Steger}}, \bibinfo {author} {\bibfnamefont {A.}~\bibnamefont {Yang}}, \bibinfo {author} {\bibfnamefont {M.~L.~W.}\ \bibnamefont {Thewalt}}, \bibinfo {author} {\bibfnamefont {M.}~\bibnamefont {Cardona}}, \bibinfo {author} {\bibfnamefont {H.}~\bibnamefont {Riemann}}, \bibinfo {author} {\bibfnamefont {N.}~\bibnamefont {Abrosimov}}, \bibinfo {author} {\bibfnamefont {M.}~\bibnamefont {Churbanov}}, \bibinfo {author} {\bibfnamefont {A.}~\bibnamefont {Gusev}}, \bibinfo {author} {\bibfnamefont {A.}~\bibnamefont {Bulanov}}, \bibinfo {author} {\bibfnamefont {I.}~\bibnamefont {Kovalev}}, \bibinfo {author} {\bibfnamefont {A.}~\bibnamefont {Kaliteevskii}}, \bibinfo {author} {\bibfnamefont {O.}~\bibnamefont {Godisov}}, \bibinfo {author} {\bibfnamefont {P.}~\bibnamefont {Becker}}, \bibinfo {author} {\bibfnamefont {H.~J.}\ \bibnamefont {Pohl}}, \bibinfo {author} {\bibfnamefont {E.}~\bibnamefont {Haller}}, \ and\ \bibinfo {author} {\bibfnamefont
  {J.}~\bibnamefont {Ager}},\ }\href {\doibase 10.1103/PhysRevB.80.115204} {\bibfield  {journal} {\bibinfo  {journal} {Phys. Rev. B.}\ }\textbf {\bibinfo {volume} {80}},\ \bibinfo {pages} {115204} (\bibinfo {year} {2009})}\BibitemShut {NoStop}%
\bibitem [{\citenamefont {Swartz}\ \emph {et~al.}(1980)\citenamefont {Swartz}, \citenamefont {Lemmon},\ and\ \citenamefont {Thomas}}]{Swartz1980OpticalSilicon}%
  \BibitemOpen
  \bibfield  {author} {\bibinfo {author} {\bibfnamefont {J.~C.}\ \bibnamefont {Swartz}}, \bibinfo {author} {\bibfnamefont {D.~H.}\ \bibnamefont {Lemmon}}, \ and\ \bibinfo {author} {\bibfnamefont {R.~N.}\ \bibnamefont {Thomas}},\ }\href {\doibase 10.1016/0038-1098(80)90065-4} {\bibfield  {journal} {\bibinfo  {journal} {Solid State Communications}\ }\textbf {\bibinfo {volume} {36}},\ \bibinfo {pages} {331} (\bibinfo {year} {1980})}\BibitemShut {NoStop}%
\bibitem [{\citenamefont {Abraham}\ \emph {et~al.}(2018)\citenamefont {Abraham}, \citenamefont {Deabreu}, \citenamefont {Morse}, \citenamefont {Shuman}, \citenamefont {Portsel}, \citenamefont {Lodygin}, \citenamefont {Astrov}, \citenamefont {Abrosimov}, \citenamefont {Pavlov}, \citenamefont {H{\"{u}}bers}, \citenamefont {Simmons},\ and\ \citenamefont {Thewalt}}]{Abraham2018FurtherSilicon}%
  \BibitemOpen
  \bibfield  {author} {\bibinfo {author} {\bibfnamefont {R.~J.}\ \bibnamefont {Abraham}}, \bibinfo {author} {\bibfnamefont {A.}~\bibnamefont {Deabreu}}, \bibinfo {author} {\bibfnamefont {K.~J.}\ \bibnamefont {Morse}}, \bibinfo {author} {\bibfnamefont {V.~B.}\ \bibnamefont {Shuman}}, \bibinfo {author} {\bibfnamefont {L.~M.}\ \bibnamefont {Portsel}}, \bibinfo {author} {\bibfnamefont {A.~N.}\ \bibnamefont {Lodygin}}, \bibinfo {author} {\bibfnamefont {Y.~A.}\ \bibnamefont {Astrov}}, \bibinfo {author} {\bibfnamefont {N.~V.}\ \bibnamefont {Abrosimov}}, \bibinfo {author} {\bibfnamefont {S.~G.}\ \bibnamefont {Pavlov}}, \bibinfo {author} {\bibfnamefont {H.~W.}\ \bibnamefont {H{\"{u}}bers}}, \bibinfo {author} {\bibfnamefont {S.}~\bibnamefont {Simmons}}, \ and\ \bibinfo {author} {\bibfnamefont {M.~L.}\ \bibnamefont {Thewalt}},\ }\href {\doibase 10.1103/PhysRevB.98.045202} {\bibfield  {journal} {\bibinfo  {journal} {Phys. Rev. B}\ }\textbf {\bibinfo {volume} {98}},\ \bibinfo {pages} {045202} (\bibinfo {year}
  {2018})}\BibitemShut {NoStop}%
\bibitem [{\citenamefont {Pavlov}\ \emph {et~al.}(2019)\citenamefont {Pavlov}, \citenamefont {Abrosimov}, \citenamefont {Shuman}, \citenamefont {Portsel}, \citenamefont {Lodygin}, \citenamefont {Astrov}, \citenamefont {Zhukavin}, \citenamefont {Shastin}, \citenamefont {Irmscher}, \citenamefont {Pohl},\ and\ \citenamefont {H{\"{u}}bers}}]{Pavlov2019}%
  \BibitemOpen
  \bibfield  {author} {\bibinfo {author} {\bibfnamefont {S.~G.}\ \bibnamefont {Pavlov}}, \bibinfo {author} {\bibfnamefont {N.~V.}\ \bibnamefont {Abrosimov}}, \bibinfo {author} {\bibfnamefont {V.~B.}\ \bibnamefont {Shuman}}, \bibinfo {author} {\bibfnamefont {L.}~\bibnamefont {Portsel}}, \bibinfo {author} {\bibfnamefont {A.~N.}\ \bibnamefont {Lodygin}}, \bibinfo {author} {\bibfnamefont {Y.~A.}\ \bibnamefont {Astrov}}, \bibinfo {author} {\bibfnamefont {R.~K.}\ \bibnamefont {Zhukavin}}, \bibinfo {author} {\bibfnamefont {V.~N.}\ \bibnamefont {Shastin}}, \bibinfo {author} {\bibfnamefont {K.}~\bibnamefont {Irmscher}}, \bibinfo {author} {\bibfnamefont {A.}~\bibnamefont {Pohl}}, \ and\ \bibinfo {author} {\bibfnamefont {H.~W.}\ \bibnamefont {H{\"{u}}bers}},\ }\href {\doibase 10.1002/pssb.201800514} {\bibfield  {journal} {\bibinfo  {journal} {Physica Status Solidi (B) Basic Research}\ }\textbf {\bibinfo {volume} {256}} (\bibinfo {year} {2019}),\ 10.1002/pssb.201800514}\BibitemShut {NoStop}%
\bibitem [{\citenamefont {Thilderkvist}\ \emph {et~al.}(1994)\citenamefont {Thilderkvist}, \citenamefont {Kleverman},\ and\ \citenamefont {Grimmeiss}}]{Thilderkvist1994InterstitialSilicon}%
  \BibitemOpen
  \bibfield  {author} {\bibinfo {author} {\bibfnamefont {A.}~\bibnamefont {Thilderkvist}}, \bibinfo {author} {\bibfnamefont {M.}~\bibnamefont {Kleverman}}, \ and\ \bibinfo {author} {\bibfnamefont {H.~G.}\ \bibnamefont {Grimmeiss}},\ }\href {\doibase https://doi.org/10.1103/PhysRevB.49.16338} {\bibfield  {journal} {\bibinfo  {journal} {Phys. Rev. B}\ }\textbf {\bibinfo {volume} {49}},\ \bibinfo {pages} {16338} (\bibinfo {year} {1994})}\BibitemShut {NoStop}%
\bibitem [{\citenamefont {Brower}(1970)}]{Brower1970ElectronSilicon}%
  \BibitemOpen
  \bibfield  {author} {\bibinfo {author} {\bibfnamefont {K.~L.}\ \bibnamefont {Brower}},\ }\href {\doibase 10.1103/PhysRevB.1.1908} {\bibfield  {journal} {\bibinfo  {journal} {Physical Review B}\ }\textbf {\bibinfo {volume} {1}},\ \bibinfo {pages} {1908} (\bibinfo {year} {1970})}\BibitemShut {NoStop}%
\bibitem [{\citenamefont {Peale}\ \emph {et~al.}(1988)\citenamefont {Peale}, \citenamefont {Muro}, \citenamefont {Sievers},\ and\ \citenamefont {Ham}}]{Peale1988ZeemanSilicon}%
  \BibitemOpen
  \bibfield  {author} {\bibinfo {author} {\bibfnamefont {R.~E.}\ \bibnamefont {Peale}}, \bibinfo {author} {\bibfnamefont {K.}~\bibnamefont {Muro}}, \bibinfo {author} {\bibfnamefont {A.~J.}\ \bibnamefont {Sievers}}, \ and\ \bibinfo {author} {\bibfnamefont {F.~S.}\ \bibnamefont {Ham}},\ }\href {\doibase 10.1103/PhysRevB.37.10829} {\bibfield  {journal} {\bibinfo  {journal} {Phys. Rev. B}\ }\textbf {\bibinfo {volume} {37}},\ \bibinfo {pages} {10829} (\bibinfo {year} {1988})}\BibitemShut {NoStop}%
\bibitem [{Sup()}]{Supp_Mat}%
  \BibitemOpen
  \href@noop {} {}\bibinfo {note} {See Supplemental Material at {URL will be inserted by the publisher} for the donor absorption series, Zeeman eigenergies as a function of magnetic field, and the high resolution hyperfine splitting at 0 Field at 1.2K with citations \cite{Latushko1989IRSilicon, Niklas1985SILICON, Bergman1988, Gullans2015OpticalSilicon, Castner1967OrbachSilicon, Peale1988ZeemanSilicon, Abragam1951TheoryCrystals, Abragam_and_Bleaney}}\BibitemShut {NoStop}%
\bibitem [{\citenamefont {De~Tomas}\ \emph {et~al.}(2014)\citenamefont {De~Tomas}, \citenamefont {Cantarero}, \citenamefont {Lopeandia},\ and\ \citenamefont {Alvarez}}]{DeTomas2014ThermalModel}%
  \BibitemOpen
  \bibfield  {author} {\bibinfo {author} {\bibfnamefont {C.}~\bibnamefont {De~Tomas}}, \bibinfo {author} {\bibfnamefont {A.}~\bibnamefont {Cantarero}}, \bibinfo {author} {\bibfnamefont {A.~F.}\ \bibnamefont {Lopeandia}}, \ and\ \bibinfo {author} {\bibfnamefont {F.~X.}\ \bibnamefont {Alvarez}},\ }\href {\doibase 10.1098/rspa.2014.0371} {\bibfield  {journal} {\bibinfo  {journal} {Proceedings of the Royal Society A: Mathematical, Physical and Engineering Sciences}\ }\textbf {\bibinfo {volume} {470}} (\bibinfo {year} {2014}),\ 10.1098/rspa.2014.0371}\BibitemShut {NoStop}%
\bibitem [{\citenamefont {Abragam}\ and\ \citenamefont {Pryce}(1951)}]{Abragam1951TheoryCrystals}%
  \BibitemOpen
  \bibfield  {author} {\bibinfo {author} {\bibfnamefont {A.}~\bibnamefont {Abragam}}\ and\ \bibinfo {author} {\bibfnamefont {M.~H.~L.}\ \bibnamefont {Pryce}},\ }\href {\doibase 10.1098/rspa.1951.0022} {\bibfield  {journal} {\bibinfo  {journal} {Proceedings of the Royal Society of London. Series A. Mathematical and Physical Sciences}\ }\textbf {\bibinfo {volume} {205}},\ \bibinfo {pages} {135} (\bibinfo {year} {1951})}\BibitemShut {NoStop}%
\bibitem [{\citenamefont {{J. M. Luttinger}}\ and\ \citenamefont {{W. Kohn}}(1955)}]{J.M.Luttinger1955MotionFields}%
  \BibitemOpen
  \bibfield  {author} {\bibinfo {author} {\bibnamefont {{J. M. Luttinger}}}\ and\ \bibinfo {author} {\bibnamefont {{W. Kohn}}},\ }\href {\doibase 10.1103/PhysRev.97.869} {\bibfield  {journal} {\bibinfo  {journal} {Phys. Rev.}\ }\textbf {\bibinfo {volume} {97}},\ \bibinfo {pages} {869} (\bibinfo {year} {1955})}\BibitemShut {NoStop}%
\bibitem [{\citenamefont {Crouse}(2016)}]{Crouse2016OnAlgorithms}%
  \BibitemOpen
  \bibfield  {author} {\bibinfo {author} {\bibfnamefont {D.~F.}\ \bibnamefont {Crouse}},\ }\href {\doibase 10.1109/TAES.2016.140952} {\bibfield  {journal} {\bibinfo  {journal} {IEEE Transactions on Aerospace and Electronic Systems}\ }\textbf {\bibinfo {volume} {52}},\ \bibinfo {pages} {1679} (\bibinfo {year} {2016})}\BibitemShut {NoStop}%
\bibitem [{\citenamefont {ChemLin}(2026)}]{chemLin}%
  \BibitemOpen
  \bibfield  {author} {\bibinfo {author} {\bibnamefont {ChemLin}},\ }\href {https://www.chemlin.org/isotope/aluminium-27} {\enquote {\bibinfo {title} {Aluminum-27},}\ } (\bibinfo {year} {2026})\BibitemShut {NoStop}%
\bibitem [{\citenamefont {Abragam}\ and\ \citenamefont {Bleaney}(1970)}]{Abragam_and_Bleaney}%
  \BibitemOpen
  \bibfield  {author} {\bibinfo {author} {\bibfnamefont {A.}~\bibnamefont {Abragam}}\ and\ \bibinfo {author} {\bibfnamefont {B.}~\bibnamefont {Bleaney}},\ }\href@noop {} {\emph {\bibinfo {title} {Electron Paramagnetic Resonance of Transition Ions}}}\ (\bibinfo  {publisher} {Oxford University Press},\ \bibinfo {year} {1970})\BibitemShut {NoStop}%
\bibitem [{\citenamefont {Ivanov}\ \emph {et~al.}(2022)\citenamefont {Ivanov}, \citenamefont {Simoni}, \citenamefont {Lee}, \citenamefont {Liu}, \citenamefont {Jhuria}, \citenamefont {Redjem}, \citenamefont {Zhiyenbayev}, \citenamefont {Papapanos}, \citenamefont {Qarony}, \citenamefont {Kant{\'{e}}}, \citenamefont {Persaud}, \citenamefont {Schenkel},\ and\ \citenamefont {Tan}}]{Ivanov2022EffectLocalization}%
  \BibitemOpen
  \bibfield  {author} {\bibinfo {author} {\bibfnamefont {V.}~\bibnamefont {Ivanov}}, \bibinfo {author} {\bibfnamefont {J.}~\bibnamefont {Simoni}}, \bibinfo {author} {\bibfnamefont {Y.}~\bibnamefont {Lee}}, \bibinfo {author} {\bibfnamefont {W.}~\bibnamefont {Liu}}, \bibinfo {author} {\bibfnamefont {K.}~\bibnamefont {Jhuria}}, \bibinfo {author} {\bibfnamefont {W.}~\bibnamefont {Redjem}}, \bibinfo {author} {\bibfnamefont {Y.}~\bibnamefont {Zhiyenbayev}}, \bibinfo {author} {\bibfnamefont {C.}~\bibnamefont {Papapanos}}, \bibinfo {author} {\bibfnamefont {W.}~\bibnamefont {Qarony}}, \bibinfo {author} {\bibfnamefont {B.}~\bibnamefont {Kant{\'{e}}}}, \bibinfo {author} {\bibfnamefont {A.}~\bibnamefont {Persaud}}, \bibinfo {author} {\bibfnamefont {T.}~\bibnamefont {Schenkel}}, \ and\ \bibinfo {author} {\bibfnamefont {L.~Z.}\ \bibnamefont {Tan}},\ }\href {\doibase 10.1103/PhysRevB.106.134107} {\bibfield  {journal} {\bibinfo  {journal} {Phys. Rev. B}\ }\textbf {\bibinfo {volume} {106}},\ \bibinfo {pages} {134107}
  (\bibinfo {year} {2022})}\BibitemShut {NoStop}%
\bibitem [{\citenamefont {Barrett}\ and\ \citenamefont {Kok}(2005)}]{Barrett2005}%
  \BibitemOpen
  \bibfield  {author} {\bibinfo {author} {\bibfnamefont {S.~D.}\ \bibnamefont {Barrett}}\ and\ \bibinfo {author} {\bibfnamefont {P.}~\bibnamefont {Kok}},\ }\href {\doibase 10.1103/PhysRevA.71.060310} {\bibfield  {journal} {\bibinfo  {journal} {Phys. Rev. A}\ }\textbf {\bibinfo {volume} {71}},\ \bibinfo {pages} {060310} (\bibinfo {year} {2005})}\BibitemShut {NoStop}%
\bibitem [{\citenamefont {Merkel}\ \emph {et~al.}(2020)\citenamefont {Merkel}, \citenamefont {Ulanowski},\ and\ \citenamefont {Reiserer}}]{Merkel2020CoherentPurcell}%
  \BibitemOpen
  \bibfield  {author} {\bibinfo {author} {\bibfnamefont {B.}~\bibnamefont {Merkel}}, \bibinfo {author} {\bibfnamefont {A.}~\bibnamefont {Ulanowski}}, \ and\ \bibinfo {author} {\bibfnamefont {A.}~\bibnamefont {Reiserer}},\ }\href {\doibase 10.1103/PhysRevX.10.041025} {\bibfield  {journal} {\bibinfo  {journal} {Physical Review X}\ }\textbf {\bibinfo {volume} {10}},\ \bibinfo {pages} {041025} (\bibinfo {year} {2020})}\BibitemShut {NoStop}%
\bibitem [{\citenamefont {Baron}\ \emph {et~al.}(1969)\citenamefont {Baron}, \citenamefont {Shifrin}, \citenamefont {Marsh},\ and\ \citenamefont {Mayer}}]{Baron1969ElectricalSilicon}%
  \BibitemOpen
  \bibfield  {author} {\bibinfo {author} {\bibfnamefont {R.}~\bibnamefont {Baron}}, \bibinfo {author} {\bibfnamefont {G.~A.}\ \bibnamefont {Shifrin}}, \bibinfo {author} {\bibfnamefont {O.~J.}\ \bibnamefont {Marsh}}, \ and\ \bibinfo {author} {\bibfnamefont {J.~W.}\ \bibnamefont {Mayer}},\ }\href {\doibase 10.1063/1.1658260} {\bibfield  {journal} {\bibinfo  {journal} {Journal of Applied Physics}\ }\textbf {\bibinfo {volume} {40}},\ \bibinfo {pages} {3702} (\bibinfo {year} {1969})}\BibitemShut {NoStop}%
\bibitem [{\citenamefont {Niklas}\ \emph {et~al.}(1985)\citenamefont {Niklas}, \citenamefont {Spaeth},\ and\ \citenamefont {Watkins}}]{Niklas1985SILICON}%
  \BibitemOpen
  \bibfield  {author} {\bibinfo {author} {\bibfnamefont {J.}~\bibnamefont {Niklas}}, \bibinfo {author} {\bibfnamefont {J.-M.}\ \bibnamefont {Spaeth}}, \ and\ \bibinfo {author} {\bibfnamefont {G.}~\bibnamefont {Watkins}},\ }\href {\doibase 10.1557/proc-46-237} {\bibfield  {journal} {\bibinfo  {journal} {MRS Proceedings}\ }\textbf {\bibinfo {volume} {46}} (\bibinfo {year} {1985}),\ 10.1557/proc-46-237}\BibitemShut {NoStop}%
\bibitem [{\citenamefont {Gullans}\ and\ \citenamefont {Taylor}(2015)}]{Gullans2015OpticalSilicon}%
  \BibitemOpen
  \bibfield  {author} {\bibinfo {author} {\bibfnamefont {M.~J.}\ \bibnamefont {Gullans}}\ and\ \bibinfo {author} {\bibfnamefont {J.~M.}\ \bibnamefont {Taylor}},\ }\href {\doibase 10.1103/PhysRevB.92.195411} {\bibfield  {journal} {\bibinfo  {journal} {Phys. Rev B}\ }\textbf {\bibinfo {volume} {92}} (\bibinfo {year} {2015}),\ 10.1103/PhysRevB.92.195411}\BibitemShut {NoStop}%
\bibitem [{\citenamefont {Castner}(1967)}]{Castner1967OrbachSilicon}%
  \BibitemOpen
  \bibfield  {author} {\bibinfo {author} {\bibfnamefont {T.~G.}\ \bibnamefont {Castner}},\ }\href {\doibase 10.1103/PhysRev.155.816} {\bibfield  {journal} {\bibinfo  {journal} {Phys. Rev.}\ }\textbf {\bibinfo {volume} {155}},\ \bibinfo {pages} {816} (\bibinfo {year} {1967})}\BibitemShut {NoStop}%
\bibitem [{\citenamefont {Drever}\ \emph {et~al.}(1983)\citenamefont {Drever}, \citenamefont {Hall}, \citenamefont {Kowalski}, \citenamefont {Hough}, \citenamefont {Ford}, \citenamefont {Munley},\ and\ \citenamefont {Ward}}]{Drever1983}%
  \BibitemOpen
  \bibfield  {author} {\bibinfo {author} {\bibfnamefont {R.~W.~P.}\ \bibnamefont {Drever}}, \bibinfo {author} {\bibfnamefont {J.~L.}\ \bibnamefont {Hall}}, \bibinfo {author} {\bibfnamefont {F.~V.}\ \bibnamefont {Kowalski}}, \bibinfo {author} {\bibfnamefont {J.}~\bibnamefont {Hough}}, \bibinfo {author} {\bibfnamefont {G.~M.}\ \bibnamefont {Ford}}, \bibinfo {author} {\bibfnamefont {A.~J.}\ \bibnamefont {Munley}}, \ and\ \bibinfo {author} {\bibfnamefont {H.}~\bibnamefont {Ward}},\ }\href {\doibase 10.1007/BF00702605} {\bibfield  {journal} {\bibinfo  {journal} {Applied Physics B: Lasers and Optics}\ }\textbf {\bibinfo {volume} {31}},\ \bibinfo {pages} {97} (\bibinfo {year} {1983})}\BibitemShut {NoStop}%
\bibitem [{\citenamefont {{Abraham R.J.S.}}\ \emph {et~al.}(2022)\citenamefont {{Abraham R.J.S.}}, \citenamefont {{Shuman V.B.}}, \citenamefont {{Portsel L.M.}}, \citenamefont {{Lodygin A.N.}}, \citenamefont {{Astrov Yu.A.}}, \citenamefont {{Abrosimov N.V.}}, \citenamefont {{Pavlov S. G.}}, \citenamefont {{H{\"{u}}bers H.-W.}}, \citenamefont {{Simmons S.}},\ and\ \citenamefont {{Thewalt M.L.W.}}}]{AbrahamR.J.S.2022ThermalSilicon}%
  \BibitemOpen
  \bibfield  {author} {\bibinfo {author} {\bibnamefont {{Abraham R.J.S.}}}, \bibinfo {author} {\bibnamefont {{Shuman V.B.}}}, \bibinfo {author} {\bibnamefont {{Portsel L.M.}}}, \bibinfo {author} {\bibnamefont {{Lodygin A.N.}}}, \bibinfo {author} {\bibnamefont {{Astrov Yu.A.}}}, \bibinfo {author} {\bibnamefont {{Abrosimov N.V.}}}, \bibinfo {author} {\bibnamefont {{Pavlov S. G.}}}, \bibinfo {author} {\bibnamefont {{H{\"{u}}bers H.-W.}}}, \bibinfo {author} {\bibnamefont {{Simmons S.}}}, \ and\ \bibinfo {author} {\bibnamefont {{Thewalt M.L.W.}}},\ }\href {\doibase 10.21883/sc.2022.01.53120.9583a} {\bibfield  {journal} {\bibinfo  {journal} {Semiconductors}\ }\textbf {\bibinfo {volume} {56}},\ \bibinfo {pages} {59} (\bibinfo {year} {2022})}\BibitemShut {NoStop}%
\bibitem [{\citenamefont {Altarelli}(1983)}]{Altarelli1983ExcitedSilicon}%
  \BibitemOpen
  \bibfield  {author} {\bibinfo {author} {\bibfnamefont {M.}~\bibnamefont {Altarelli}},\ }\href {\doibase 10.1016/0378-4363(83)90459-X} {\bibfield  {journal} {\bibinfo  {journal} {Physica B+C}\ }\textbf {\bibinfo {volume} {117-118}},\ \bibinfo {pages} {122} (\bibinfo {year} {1983})}\BibitemShut {NoStop}%
\bibitem [{\citenamefont {Cardona}\ \emph {et~al.}(2004)\citenamefont {Cardona}, \citenamefont {Meyer},\ and\ \citenamefont {Thewalt}}]{Cardona2004TemperatureLimit}%
  \BibitemOpen
  \bibfield  {author} {\bibinfo {author} {\bibfnamefont {M.}~\bibnamefont {Cardona}}, \bibinfo {author} {\bibfnamefont {T.~A.}\ \bibnamefont {Meyer}}, \ and\ \bibinfo {author} {\bibfnamefont {M.~L.}\ \bibnamefont {Thewalt}},\ }\href {\doibase 10.1103/PhysRevLett.92.196403} {\bibfield  {journal} {\bibinfo  {journal} {Physical Review Letters}\ }\textbf {\bibinfo {volume} {92}},\ \bibinfo {pages} {1} (\bibinfo {year} {2004})}\BibitemShut {NoStop}%
\end{thebibliography}

%

\clearpage
\onecolumngrid

\twocolumngrid
\renewcommand{\thesection}{A\arabic{section}}
\numberwithin{figure}{section}
\numberwithin{equation}{section}
\setcounter{section}{0}
\setcounter{figure}{0}

\section{Sample Preparation}
\label{sec: appendix sample prep}
We prepared two samples of aluminum-doped, isotopically enriched \TwoEightSi{} using two different float zone crystals. The first is a \qty{5}{mm} diameter cylindrical sample cut to \qty{1}{cm} in length with a doping concentration of $2.9 \times 10^{16}$\,cm$^{-3}$. It was etched in HF and HNO$_3$ to remove any surface damage before being sent to Iotron Industries for electron-irradiation at \qty{10}{MeV}. Although this sample had strong \Aliplus{} transitions, the linewidths were limited by random strain fields, particularity of the \triplet{} transition, likely produced by growth striations. It was used throughout our experiments except for the hyperfine studies. Our second sample was used to study the hyperfine interaction within the triplet state and showed small amounts of \Als{} and \Ali. This sample was a \qty{4}{\mm}$\times$\qty{4}{\mm}$\times$\qty{1}{\mm} rectangle. Both samples received 4 irradiation doses, each of 32\,kGy.

\section{Experimental Methods} \label{sec: appendix methods}
All spectra were measured using a Bruker IFS 125HR Fourier Transform Interferometer (FTIR) with various sources and settings described below. 

\textit{Zero Phonon Line} --
Resonant PL was performed by driving the \ground\,$\Rightarrow$\,2p$_\pm$ transition (1309.8\,nm) with a Toptica Continuously Tunable Laser (CTL). A tilted 1300/25\,mm band-pass filter was placed within the excitation path to remove spurious laser emission while a \qty{1400}{\nm} long-pass filter was placed in the collection path to remove the laser from the detected signal. Luminescence was detected with a liquid nitrogen-cooled germanium detector. The spectra were normalized by the instrumental response function.

\textit{Magnetospectroscopy} -- 
A superconducting split-pair magnet in a Janis cryostat was used to measure the electron Zeeman and spin-orbit interaction. The coil current was set by a CS4-10V power supply from Cryomagnetics Incorporated. A \qty{1030}{\nm} Aerodiode laser was used to excite the ensemble above the silicon band gap. A tilted \qty{1050}{\nm} band-pass cleanup filter was placed at the laser output. A \qty{1100}{\nm} long-pass filter was placed before the entrance to a 3.0\,mm core diameter 77634 Newport liquid light-guide to transport the luminescence to external port of the FTIR. The resolution was 0.1\,cm$\mathrm{^{-1}}$.

The hyperfine structure was measured in a separate cryostat using a Toptica CTL resonant on the \ground$\Rightarrow\mathrm{2p_\pm}$ transition. The sample was inserted into a brass pocket lined with gold foil to enhance the collected luminescence. The \qty{0}{\mT} spectrum was measured with a resolution set to 0.002\,cm$\mathrm{^{-1}}$. The sample and pocket were then inserted into a permanent magnet assembly to measure a spectrum at \qty{109.9(0.1)}{mT}. The resolution was set to 0.003\,cm$\mathrm{^{-1}}$.

\textit{Temperature Dependent PL} --
Spectra were collected over a range of temperatures using an above-band IRCL-500-1047 CristaLaser for excitation. The temperature of the sample was set between 4.2 and 180\,K using a Lakeshore 335 temperature controller. The resolution of the FTIR was set to 0.5\,cm$\mathrm{^{-1}}$.

\textit{Pulsed Luminescence} -- 
The excited state lifetimes were measured using pulsed excitation and time resolved single photon counting. An ID-230 InGaAs avalanche photodiode from IDQ was configured with a dead-time of \qty{2}{\micro s}. Sample temperatures were maintained using a Lakeshore temperature controller. The pulse sequence was generated using a DG 645 delay generator from Stanford Research Systems, and a Swabian time tagger binned single photon events within the measurement window set by the delay generator. 

Using both above-band and resonant excitation, two time-resolved PL datasets were measured. A Toptica DL100 diode laser, stabilized to an external cavity from Stable Laser Systems using the Pound-Drever-Hall technique \cite{Drever1983}, resonantly pulsed the \ground{} $ \Rightarrow \mathrm{2p_0}$ transition. An Aerodiode optical amplifier pulsed the excitation intensity, and a \qty{1350}{\nm} band-pass cleanup filter removed spurious emission from the amplifier. PL was transmitted through three \qty{1400}{\nm} long-pass filters. A second data set was generated using a pulsed \qty{1030}{\nm} Aerodiode laser providing above-band excitation. In this case, the fluorescence from other defects was reduced using a \qty{1319}{\nm} long-pass filter.

\textit{Absorption} -- 
All absorption measurements were performed using a tungsten-halogen lamp transmitted through the \qty{1}{cm} length of the large cylindrical sample and detected on a liquid nitrogen cooled InSb detector. 

\section{Excited State Lifetimes} \label{sec: appendix lifetimes}
\begin{figure}[b]
    \centering
    \includegraphics[width = 8.5cm]{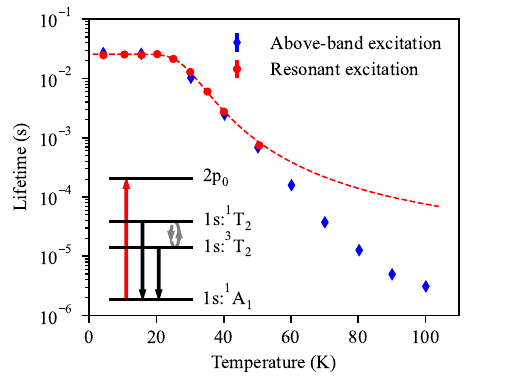}
    \caption{Excited state lifetime as a function of temperature. (Blue) Above-band excitation. (Red) resonant excitation of  $\mathrm{2p_0}$. (Inset) level structure showing decay from the singlet and triplet levels (black) and thermally-activated mixing (grey). A Boltzmann model (dashed red) is fitted to the data below \qty{50}{\kelvin}.} 
    \label{fig: lifetime distributions}
\end{figure}

The temperature-dependant excited state decay rate $\Gamma$ is a weighted sum of the singlet and triplet rates $\Gamma_\mathrm{s, t}$ with weights given by the Boltzmann distribution
\begin{equation}
    \Gamma = \frac{\mathrm{g_t}\Gamma_\mathrm{t} + \left(\mathrm{g_s}\Gamma_\mathrm{s}\right)\exp\left(-\Delta E/k_\mathrm{B}T\right)}{\mathrm{g_t} + \mathrm{g_s}\exp\left(-\Delta E/k_\mathrm{B} T\right)} \,,
    \label{eq: rate equation}
\end{equation}
where $\mathrm{g_{s(t)}}$ represent the degeneracy factors of 1(3) of the singlet (triplet) states. For fitting, the singlet-triplet splitting is fixed to $\Delta E =$ \qty{21.26}{\meV} using the measured energies of \triplet$_{(\mathrm{\mathcal{J}=1)}}$\,$\Rightarrow$\ground{} and \ground{}$\Rightarrow$\singlet{} from \cref{table: ali levels}.

For a range of temperatures, up to \qty{100}{K}, the total decay rate was extracted using a least-squares single-exponential fit to the PL decay data. Its temperature dependence, along with a fit between \qtyrange{4.2}{50}{\K} to \cref{eq: rate equation}, is shown in \Cref{fig: lifetime distributions}. We fit a singlet and triplet lifetime of $\tau_\mathrm{s} = \qty{1.97(0.04)}{\us}$ and $\tau_\mathrm{t} = \qty{25.5(0.5)}{\ms}$. Above \qty{50}{K}, the measured lifetimes deviate from the fit, suggesting that system begins to thermalize into additional states or ionize. 

\section{Temperature-Dependent Photoluminescence} \label{sec: appendix intensity ratio}
\begin{figure}[t]
    \centering
    \includegraphics[width=8.5cm]{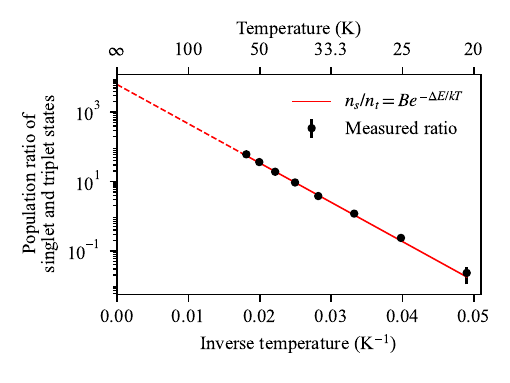}
    \caption{The log ratio between the integrated singlet and triplet intensities as a function of 1/T is shown.}
    \label{fig: arrhenius}
\end{figure}

\Cref{fig: arrhenius} displays the integrated intensity ratio of the singlet and triplet states as a function of temperature. The integrated intensities are found by fitting two Gaussian-Lorentzian product functions to each PL spectrum. We denote the population of the singlet (triplet) as $\mathrm{n_{s(t)}}$, and the pre-exponential factor ($B$) describes the system in the high-temperature limit. Using a linear fit to the Arrhenius equation in \cref{eq: arrhenius}, we find an activation energy ($\Delta E$) of \qty{21.53(0.18)}{meV}, in agreement with the measured energy difference between the two states.
\begin{equation}
    \ln{\left(\frac{n_\mathrm{s}}{n_\mathrm{t}}\right)} = \ln{B}-\frac{\Delta E}{k_\mathrm{B}}\left( \frac{1}{T}\right)
    \label{eq: arrhenius}
\end{equation}

In thermal equilibrium, the population transfer between the two states is constrained by $n_\mathrm{s} \frac{\tau_\mathrm{s}}{\mathrm{g_s}} = n_\mathrm{t}\frac{\tau_\mathrm{t}}{\mathrm{g_t}}$. The population ratio was taken at the infinite temperature limit of \cref{eq: arrhenius}, and the pre-exponential factor was used to predict the singlet lifetime in \cref{eq: arrhenius lifetime}. 
\begin{equation}
    B = \lim_{T\Rightarrow\infty} \left(\frac{n_\mathrm{s}}{n_\mathrm{t}}\right)=\frac{\mathrm{g_s}\Gamma_\mathrm{s}}{\mathrm{g_t}\Gamma_\mathrm{t}} = \frac{\mathrm{g_s}\tau_\mathrm{t}}{\mathrm{g_t} \tau_\mathrm{s}}
    \label{eq: arrhenius lifetime}
\end{equation}
With the triplet state lifetime from \cref{sec: appendix lifetimes}, we estimate a \stateT{} singlet to ground lifetime of \qty{1.74(0.4)}{\us}, differing by $5\sigma$ from the independent determination of $\tau_s$ by lifetime measurements. Thermalization into an additional dark state, such as the forbidden \stateE{} exchange-split states, may be responsible \cite{AbrahamR.J.S.2022ThermalSilicon, Altarelli1983ExcitedSilicon}. Considering the precision of these two measurements, we report $\tau_s=\qty{1.8(0.2)}{\us}$. 

\section{AM1 Transition Energies}
\label{sec: appendix AM1 table}
We present the transition energies of the \Aliplus{} series. The \triplet{} sublevels were determined in \cref{sec: excited state} using above-band PL. The rest were measured in absorption (SM \cref{sec: sub mat absorption} \cite{Supp_Mat}).
\begin{table}[ht]
    \centering
    \renewcommand*\arraystretch{1.3}
    \centering 
    \begin{tabular*}{\columnwidth}{@{\extracolsep{\fill}}llcc}
        \hline \hline
        \multirow{2}{*}{Series} &\multirow{2}{*}{Label} &\multicolumn{2}{c}{Transition Energy (meV)}\\
        & &Previous\,\cite{Latushko1989IRSilicon} &This Work\\
        \hline
        \multirow{15}{*}{AM1}
        &\triplet$_{(\mathcal{J}=0)}$\,$\Rightarrow$\ground         &-        &774.83(1)\\
        &\triplet$_{(\mathcal{J}=1)}$\,$\Rightarrow$\ground         &774.8(1) &774.87(1)\\
        &\triplet$_{(\mathcal{J}=2)}$\,$\Rightarrow$\ground         &-        &774.96(1)\\
        \cline{2-4} 
        \noalign{\vskip\doublerulesep \vskip-\arrayrulewidth} \cline{2-4}
        &\ground$\Rightarrow$\singlet           &796.4(1) &796.14(1)\\
        &\ground$\Rightarrow$ 2s:\ce{T_2}       &-        &931.65(2)\\
        &\ground$\Rightarrow \mathrm{2p_0}$     &924.6(3) &924.48(1)\\ 
        \cline{2-4} 
        \noalign{\vskip\doublerulesep \vskip-\arrayrulewidth} \cline{2-4}
        &\multirow{2}{*}{\ground$ \Rightarrow \mathrm{2p_\pm}$}      
                                                        &946.8(3) &946.60(1),\\
        &                                               &-        &946.87(1)\\ 
        \cline{2-4} 
        \noalign{\vskip\doublerulesep \vskip-\arrayrulewidth} \cline{2-4}
        &\ground$ \Rightarrow \mathrm{3p_0}$    &950.0(3) &949.95(1)  \\
        &\ground$ \Rightarrow \mathrm{3d_0}$    &-        &958.00(1)  \\
        &\ground$ \Rightarrow \mathrm{3p_\pm}$  &960.6(5) &960.50(1)  \\
        &\ground$ \Rightarrow \mathrm{4p_0}$    &959.4(3) &959.34(1)  \\
        &\ground$ \Rightarrow \mathrm{4p_\pm}$  &963.5(5) &963.86(1)  \\
        &\ground$ \Rightarrow \mathrm{5p_0}$    &963.5(5) &964.43(1)  \\
        &\ground$ \Rightarrow \mathrm{5p_\pm}$  &-        &966.45(1)  \\
        &\ground$ \Rightarrow \mathrm{6p_0}$    &-        &965.66(1)  \\
        &\ground$ \Rightarrow \mathrm{6p_\pm}$  &-        &967.34(1)  \\
        \hline \hline
    \end{tabular*}
    \caption{The transition energies of the AM1 series.}
    \label{table: ali levels}
    \vspace{-5mm}
\end{table} 

\clearpage 
\renewcommand{\thesection}{S\arabic{section}}
\renewcommand{\theequation}{S\arabic{section}.\arabic{equation}}
\renewcommand{\thefigure}{S\arabic{section}.\arabic{figure}}
\renewcommand{\thetable}{S\arabic{section}.\arabic{table}}
\setcounter{section}{0}
\setcounter{figure}{0}
\setcounter{equation}{0}
\setcounter{table}{0}

\title{Supplementary information: Optically Resolved Excited State Hyperfine Structure of a Silicon Colour Centre in the Telecom Bands}
\maketitle

\onecolumngrid 

\section{Absorption Spectrum of the Donor Series} \label{sec: sub mat absorption}
\begin{figure}[ht]
    \centering
    \includegraphics{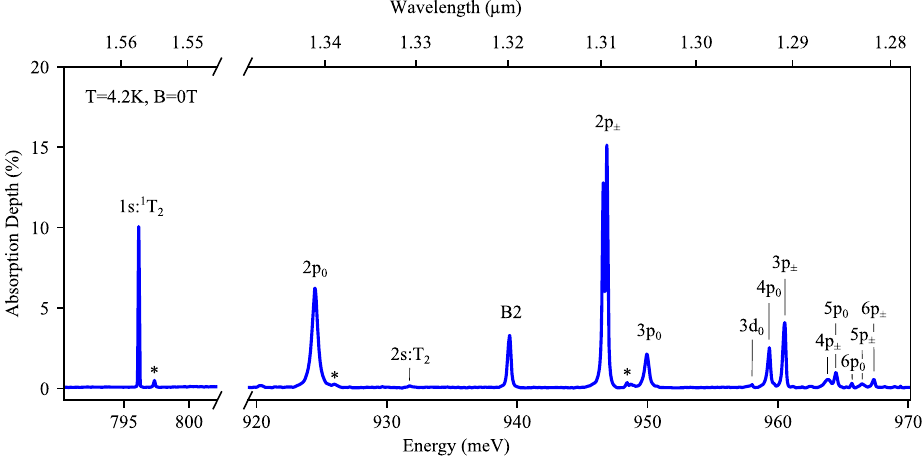}
    \caption{The absorption spectrum of the \qty{1}{cm} long \TwoEightSi sample after electron irradiation and before any thermal annealing.  It is dominated by transitions to the excited states of the AM1 series, although very weak features associated with the AM2 series, indicated by asterisks, are visible.}
    \label{fig: SM AM1 absorption}
\end{figure}

The absorption spectrum of electron-irradiated, isotopically purified \TwoEightSi{} doped with aluminum is shown in \cref{fig: SM AM1 absorption}, revealing a strong singly-ionized donor series consistent with Latushko \textit{et al.}~\cite{Latushko1989IRSilicon}. Following their nomenclature, we label this series AM1 and also observe the correlated B2 transition. This absorption feature was believed to arise from a multi-aluminum center; however, we propose that it is a transition of the deep \Aliplusplus{} charge state. The small peaks at 797.3, 926, and \qty{948.5}{meV} are consistent with the AM2 series that dominate the absorption spectrum of thermally annealed electron-irradiated aluminum-doped samples \cite{Latushko1989IRSilicon}. The optical properties of the AM2 series, along with discussions of the related B2 and B1 transitions, and the absorption series of several newly observed neutral donors, will be discussed in a forthcoming publication.

We report previously unobserved transitions of the AM1 series including $\mathrm{5p_\pm, 6p_0}$, and $\mathrm{6p_\pm}$ and a splitting in $\mathrm{2p_\pm}$ of \qty{0.27 \pm0.01}{meV} which we attribute to the internal electronic structure of the defect. Additionally, we identify the $\mathrm{2s}$:$\mathrm{T_2}$ and $\mathrm{3d_0}$ even-parity transitions. The \triplet$\Rightarrow$\ground{} is absent from the absorption spectrum due to its low oscillator strength. The transition energies of the AM1 series are listed in \cref{table: ali levels} of the main text.

\section{Zeeman and Spin-Orbit Eigenstates}
\label{sec: sup Zeeman}
Optical excitation from the $\mathrm{1s^1}$:$\mathrm{A_1}$ ground state to \singlet{} is forbidden by symmetry; however, it is permitted within central-cell corrections to EMT. Optical excitation to the \triplet{} requires an effective total angular momentum $\mathcal{J}$ to relax the spin selection criteria through mixing with \singlet. These total angular momentum states are formed from the spin-orbit interaction 
\begin{equation}
    \mathcal{H}_\mathrm{so}=\frac{\hbar^2}{2m_0^2c^2}\sum_{i=1}^2 \left(\nabla U(r)\times\mathbf{p}_i\right) \cdot \mathbf{s}_i
    \label{eq: spin-orbit full}
\end{equation}
describing the interaction of each electron's magnetic moment with their induced orbital character while they propagate through a potential. The potential $U(r)$ contains the attractive impurity potential, its central-cell corrections \cite{Bergman1988, Gullans2015OpticalSilicon}, and the tetrahedral crystal field \cite{Castner1967OrbachSilicon}.

The 1s:T$\mathrm{_2}$ state is composed of antisymmetric combinations of conduction valleys along equivalent [100] directions, resulting in a 3-fold orbital degeneracy and an effective angular momentum $\mathcal{L}=1$. This emergent operator is defined to satisfy the commutation relations of angular momentum algebra. The degeneracy of the total angular momentum, formed within a manifold of LS-coupling by $\mathbf{\mathcal{J}}=\mathbf{\mathcal{L}} + \mathbf{S}$, is lifted by the effective spin-orbit coupling into the levels $\mathcal{J}\in\{0,1,2\}$ and split from the unperturbed state by $-2\lambda$, $-\lambda$, and $\lambda$, respectively. The effective spin-orbit Hamiltonian is given by $\mathcal{H}_\mathrm{so}=\lambda (\mathbf{\mathcal{L}}\cdot\mathbf{S})$ where S is a vector of $3\times3$ spin operators for $\mathrm{S=1}$ and the angular momentum operators are defined in \cref{eq: angular momentum}. 
\begin{equation}
    \mathcal{L}_x = 
    \begin{pmatrix}
        0  &0 &0  \\
        0  &0 &-i \\
        0  &i &0  \\ 
    \end{pmatrix},
    \hspace{1.5em}
    \mathcal{L}_y = 
    \begin{pmatrix}
        0   &0 &i \\
        0   &0 &0 \\
        -i  &0 &0 \\ 
    \end{pmatrix},
    \hspace{1.5em}
    \mathcal{L}_z = 
    \begin{pmatrix}
        0  &-i &0 \\
        i  &0  &0 \\
        0  &0  &0 \\ 
    \end{pmatrix}
    \label{eq: angular momentum}
\end{equation}

The linear Zeeman interaction splits each sublevel into their $m_\mathcal{J}$ components in the presence of an external magnetic field. The Hamiltonian describing this interaction is $\mathcal{H}_\mathrm{Z} = \mu_\mathrm{B} g_\mathrm{s}(\mathbf{S}\cdot\mathbf{B}) + \mu_\mathrm{B} g_\mathrm{\mathcal{L}}(\boldsymbol{\mathcal{L}}\cdot\mathbf{B})$ where $g_\mathcal{L}$ and $g_\mathrm{s}$ denote the orbital angular momentum and spin Land\'e factors, and $\mathrm{\mu_B}$ is the Bohr magneton. For all fits in the manuscript, we fix $\mathbf{B}_0$ to coincide with the approximate field alignment along the [110] crystal direction $(\hat{x} +\hat{y})/\sqrt{2}$.

The $9\times9$ Hamiltonian is diagonal along the component of $\mathcal{J}$ parallel with the applied field. In the coupled basis, these states are labelled with the quantum numbers $\ket{S\mathcal{L}\mathcal{J} m_\mathcal{J}}$. Off-diagonal elements describe mixing between states of $\Delta\mathcal{J}=\pm 1$ or by $\Delta m_\mathcal{J}=\pm 1$, and states with different magnetic quantum numbers with respect to the component of $\mathcal{J}$ do not mix. The system is, therefore, isotropic \cite{Peale1988ZeemanSilicon}. The energies $E_{m_\mathcal{J}}$ describing states with $m_\mathcal{J}=\pm 1$ are 
\begin{subequations}
    \begin{align}
        E_1& = \frac{1}{2}(g_\mathrm{s} + g_\mathcal{L})\mu_\mathrm{B} B \pm \left[\lambda^2 +\frac{1}{4}(g_\mathrm{s} - g_\mathcal{L})^2(\mu_\mathrm{B} B)^2\right]^\frac{1}{2}
        \label{eq mj = 1} \,,
        \\
        E_{-1}& = -\frac{1}{2}(g_\mathrm{s} + g_\mathcal{L})\mu_\mathrm{B} B \pm \left[\lambda^2 +\frac{1}{4}(g_\mathrm{s} - g_\mathcal{L})^2(\mu_\mathrm{B} B)^2\right]^\frac{1}{2} \,.
        \label{eq mj = -1}
    \end{align}
\end{subequations}
\noindent The second term in \cref{eq mj = 1} and \cref{eq mj = -1} is negative for states within sublevel $\mathcal{J}=1$ and positive for $\mathcal{J}=2$. The eigenenergies describing states with $m_\mathcal{J}=\pm 2$ are
\begin{equation}
    \label{eq: mj =2}
    E_{\pm 2} = \lambda \pm (g_\mathrm{s} + g_\mathcal{L})\mu_\mathrm{B} B \,.
\end{equation}
\noindent Lastly, the three states with $m_\mathcal{J}=0$, one in each $\mathcal{J}$ sublevel, are the roots of the characteristic polynomial 
\begin{equation}
    \label{eq: poly}
    E^3 +2\lambda E^2 -\left(\lambda^2+ (g_\mathrm{s} - g_\mathcal{L})^2(\mu_\mathrm{B} B)^2\right)E - 2\lambda^3=0 \,.
\end{equation}    

\section{Zero-Field Hyperfine Spectrum} \label{sec: sup Zero field hyperfine early}
\begin{figure}[ht]
    \centering
    \includegraphics{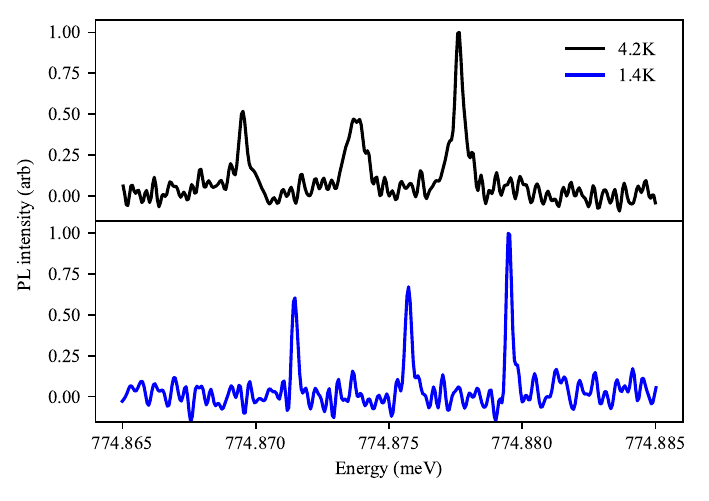}
    \caption{PL spectra of the zero-field hyperfine levels within the $\mathcal{J}=1$ sublevel of \triplet{} at \qty{4.2}{K} (black) and \qty{1.4}{K} (blue).}
     \label{fig: early Zero Field hyperfine data}
\end{figure}

Our earliest PL spectrum of the low Al concentration $^{28}$Si sample was taken at $\sim$\qty{1.4}{K} and shows narrow transition linewidth limited by the instrumental resolution of \qty{0.002}cm$^{-1}$. \Cref{fig: early Zero Field hyperfine data} contrasts this data (blue) with the spectrum used within the main text (black). Both spectra show narrow lines consistent with the $\mathcal{F}=3/2$, $5/2$, and $7/2$ features within \triplet{}$_{(\mathcal{J}=1)}$. Not only were these taken at different temperatures, but they are approximately 6 years apart. After this time, being stored at room temperature, the sample  still produced luminescence from \Aliplus{}. The two spectra are shifted by \qty{1.9\pm0.1}{\micro eV} due to the temperature dependence of the Si band gap \cite{Cardona2004TemperatureLimit}.

\end{document}